\pdfoutput=1
\documentclass[usenatbib]{mnras}

\usepackage[T1]{fontenc}
\usepackage{newtxtext,newtxmath}
\usepackage{graphicx}
\usepackage{amsmath}
\usepackage{booktabs}
\usepackage{url}
\usepackage{xspace}
\usepackage{xcolor}
\usepackage{orcidlink}

\graphicspath{{./}}

\newcommand{\kmsmpc}{\ensuremath{\,\rm km\,s^{-1}\,Mpc^{-1}}\xspace}
\newcommand{\kms}{\ensuremath{\,\rm km\,s^{-1}}\xspace}
\newcommand{\hzero}{\ensuremath{H_{0}}\xspace}
\newcommand{\dL}{\ensuremath{d_{L}}\xspace}
\newcommand{\sigmavp}{\ensuremath{\sigma_{v_p}}\xspace}
\newcommand{\IMR}{IMRPhenomD\_NRTidalv2\xspace}
\newcommand{\IMRX}{IMRPhenomXAS\_NRTidalv3\xspace}
\newcommand{\XPHM}{IMRPhenomXPHM\xspace}
\newcommand{\TF}{TaylorF2\xspace}
\newcommand{\bjns}{\textsc{blackjax-ns}\xspace}
\newcommand{\bilby}{\textsc{bilby}\xspace}
\newcommand{\pbilby}{\textsc{parallel\_bilby}\xspace}
\newcommand{\jax}{\textsc{jax}\xspace}
\newcommand{\ripple}{\textsc{ripple}\xspace}

\title[Rapid GW170817 \hzero inference]{Rapid Hubble constant inference from GW170817 using GPU-accelerated nested sampling: prior sensitivity and the limits of post-hoc reweighting}

\author[Yang, Prathaban, Yallup \& Handley]{
Ming Han Yang\,\orcidlink{0009-0002-7298-7740},$^{1}$\thanks{E-mail: mhy32@cantab.ac.uk}
Metha Prathaban,$^{2,3}$
David Yallup,$^{4,2}$
and Will Handley$^{4,2}$
\\
$^{1}$Independent researcher (formerly Institute of Astronomy and St John's College, University of Cambridge), UK\\
$^{2}$Kavli Institute for Cosmology, University of Cambridge, Madingley Road, Cambridge CB3 0HA, UK\\
$^{3}$Department of Physics, University of Cambridge, JJ Thomson Avenue, Cambridge CB3 0HE, UK\\
$^{4}$Institute of Astronomy, University of Cambridge, Madingley Road, Cambridge CB3 0HA, UK
}

\date{\today}
\pubyear{2026}

\begin{document}
\label{firstpage}
\pagerange{\pageref{firstpage}--\pageref{lastpage}}
\maketitle

\begin{abstract}
The bright-siren measurement of the Hubble constant from GW170817 \citep{Abbott2017H0} assumes that switching from a volumetric to a uniform-in-$d_L$ luminosity-distance prior can be implemented by post-hoc reweighting of the baseline samples, rather than by re-running the inference under the target prior. Using a GPU-native heterodyned nested sampling pipeline that completes the full $n_{\rm live}=5000$ analysis in $\sim 13\,\rm min$ on a single A100, we recompute the GW170817 \hzero posterior under four prior variants for the modern aligned-spin tidal waveform IMRPhenomXAS\_NRTidalv3. Switching from the volumetric to a uniform-in-$d_L$ distance prior raises the high-tail probability $P(\hzero>120\kmsmpc)$ from $0.017$ to $0.159$ when imposed during sampling and shifts the weighted-median \hzero\ from $77.6$ to $87.6\kmsmpc$, while the binned MAP stays at $70.5\kmsmpc$: both the tail and the bulk move under a change of prior that leaves the mode in place. Post-hoc reweighting of the baseline samples to the same target prior recovers only $P=0.041$ in the tail, approximately $17\,\%$ of the directly sampled shift. The three prior variants that carry an independent nested sampling evidence agree to $\Delta\ln Z\lesssim 1.8$, so the data show at most a weak preference among the distance priors; the tail and bulk shifts are therefore properties of the prior, not a data update. Targeted mode-isolated runs reveal a $(\dL,\iota)$ bimodality whose high-\hzero, low-\dL\ branch (Mode~B; $|\ln\mathcal{B}_{\rm B/A}|<1$) the volumetric prior assigns negligible mass: this is the mechanism behind the reweighting deficit. The reweighted posterior has a lower effective sample size than the baseline, independently flagging the coverage failure. The runtime budget makes full-sample prior-sensitivity reruns the default robustness tool for bright-siren cosmology, replacing post-hoc reweighting.
\end{abstract}

\begin{keywords}
gravitational waves -- methods: data analysis -- methods: statistical -- stars: neutron -- cosmological parameters -- distance scale
\end{keywords}

\section{Introduction}
\label{sec:introduction}

The present-day expansion rate of the Universe, quantified by the Hubble constant \hzero, is the subject of one of the most persistent tensions in contemporary cosmology. Measurements anchored in the early Universe, derived from observations of the Cosmic Microwave Background (CMB) assuming flat $\Lambda$CDM, favour a low value \citep{Planck2020}, while late-Universe distance ladder measurements based on Cepheids and Type Ia supernovae find a significantly higher value \citep{Riess2016,Riess2022}. The statistical and systematic discrepancy between these two families of measurement is now well established \citep{DiValentino2021HubbleReview}, and its resolution -- whether through unrecognised systematics or new cosmological physics -- motivates independent probes of the expansion rate.

Gravitational-wave standard sirens are such a probe. The waveform amplitude of a compact-binary merger fixes the luminosity distance directly, without reference to an astrophysical distance ladder \citep{Schutz1986}; \citet{PalmeseMastrogiovanni2025} review the methodology and its current status. When the merger is accompanied by an electromagnetic counterpart that identifies a host galaxy, the host redshift can be combined with the gravitational-wave distance to infer \hzero from a single event \citep{HolzHughes2005}. The binary neutron-star merger GW170817 \citep{Abbott2017GW170817Discovery} is the canonical bright siren and the only confirmed example to date. Its uniqueness has motivated complementary galaxy-catalogue (dark-siren) analyses that infer the host redshift statistically rather than from a unique electromagnetic counterpart \citep{LVK2026GWTC5Cosmology}; the same scarcity of bright sirens makes careful treatment of the one we do have especially important. GW170817's association with the kilonova AT2017gfo in NGC~4993 yielded the first gravitational-wave \hzero measurement, $\hzero=70_{-8}^{+12}\kmsmpc$ at $68\,\%$ credibility \citep{Abbott2017H0}. That measurement has since been revisited from several directions: superluminal-jet constraints on the binary inclination \citep{Mooley2018Nature,Hotokezaka2019}, improved peculiar-velocity modelling \citep{Nicolaou2020,Mukherjee2021Velocity,HowlettDavis2020}, and multi-wavelength afterglow analyses of GRB~170817A that constrain the viewing angle and shift \hzero\ towards the local distance-ladder value \citep{Wang2023GW170817H0}, most recently tightening the single-event result to $\hzero=75.5_{-5.4}^{+5.3}\kmsmpc$ at $\sim 7\,\%$ precision \citep{Palmese2024GW170817H0}. A parallel line of work has argued that residual binary-viewing-angle uncertainty contributes a $\gtrsim 10\,\%$ systematic on the GW170817 \hzero \citep{SalvareseChen2024}; the prior-sensitivity question we address here sits in the same systematics-quantification programme.

For a single bright siren, the \hzero uncertainty is dominated by the luminosity-distance--inclination degeneracy: a face-on binary at larger \dL can produce the same observed strain as a more inclined binary at smaller \dL, with the two solutions mapping to different values of $\hzero \propto (v_{\rm r}-v_{p})/\dL$ for a fixed host recessional velocity $v_{\rm r}$ and peculiar velocity $v_{p}$. The resulting posterior is broad and skewed, and only weakly informative on \hzero from one event. In such a regime, the measurement is more sensitive to prior choices than the statistical uncertainty alone would suggest, because the prior shapes the tails of the posterior rather than just its peak. The original GW170817 analysis adopted a volumetric distance prior $\pi(\dL)\propto\dL^{2}$ over $[10,75]\,\rm Mpc$ and reported that switching to a flat-in-redshift prior (numerically ${\approx}$ uniform-in-$d_L$ at this redshift; see Section~\ref{sec:priors}), implemented by \emph{reweighting} posterior samples, left the bulk of the inferred \hzero\ posterior consistent with the volumetric baseline (their Extended Data Table~1; \citealp{Abbott2017H0}). Whether the same conclusion holds when the prior is imposed during sampling rather than after the fact is the question this paper addresses.

Reweighting is attractive because it avoids rerunning expensive Bayesian inference. For a posterior $p(\theta\mid d,\pi_{0})$ obtained under baseline prior $\pi_{0}$, the target posterior under an alternative prior $\pi_{1}$ can be estimated by assigning each sample a weight proportional to $\pi_{1}(\theta)/\pi_{0}(\theta)$. This procedure is statistically consistent provided the baseline samples cover the support of the target posterior. If $\pi_{1}/\pi_{0}$ is large in regions that the baseline run has sampled sparsely, the reweighted estimator has high variance and systematically low tail mass \citep[see e.g.][]{Speagle2020Dynesty}; \citet{Ashton2025Reweighting} gives a recent guide to reweighting gravitational-wave posteriors and its importance-sampling diagnostics. For GW170817, the baseline volumetric prior places little mass at low \dL; a uniform-in-$d_L$ target up-weights exactly this region, and the validity of reweighting is not obvious a priori.

Testing reweighting against a direct posterior obtained by sampling under the target prior requires repeated full gravitational-wave parameter estimation. In CPU frameworks such as \bilby \citep{Ashton2019Bilby} and \pbilby, single-event binary-neutron-star analyses with phase-marginalised frequency-domain likelihood are sufficiently expensive that systematic exploration of distance, velocity, sampler, and waveform assumptions is rarely carried out in practice. Several recent developments make this constraint less severe: relative binning, also known as the heterodyned likelihood \citep{Cornish2010,Zackay2018RelativeBinning}, reduces the effective frequency grid for a long inspiral to a small number of bins; differentiable waveform libraries \citep{Edwards2023Ripple} implemented in the \jax\ framework \citep{Bradbury2018JAX} enable end-to-end GPU evaluation of the likelihood; and accelerated nested sampling kernels, both CPU-side via normalising-flow proposals \citep{Williams2021Nessai} and GPU-native via vectorised slice sampling \citep{BlackJAX,Yallup2026BlackjaxNS}, retain the evidence estimates that nested sampling provides for free while permitting large live-point counts and repeated full reruns. Related accelerated binary-neutron-star pipelines \citep{Krishna2023RelativeBinningBilby,Wong2023Jim,Wouters2024JimBNS} and forward-looking analyses \citep{HuVeitch2025} underscore the practical importance of compute-efficient parameter estimation for the third-generation detector era.

In this paper we use a GPU-native nested sampling pipeline -- a \bjns\ slice-sampling kernel with a \jax\ implementation of the heterodyned likelihood -- to revisit the GW170817 bright-siren \hzero inference of \citet{Abbott2017H0}. The inner kernel we adopt is the GPU-native Nested Slice Sampling kernel of \citet{Yallup2026BlackjaxNS}, implemented in the \bjns\ framework \citep{BlackJAX}. The same framework was extended by \citet{Prathaban2025GPUSpeed} to a GPU-native form of the \bilby\ ``acceptance-walk'' kernel, applied there to binary black hole events; the Nested Slice Sampling kernel itself was demonstrated on (short-duration, binary black hole) gravitational-wave signals in the brief conference note of \citet{Yallup2025BlackJAXNS}. Here we apply the Nested Slice Sampling kernel at full production scale to a long-duration binary neutron star signal. Our focus is not a new single-event measurement of \hzero, but a controlled, multi-axis assessment of how that measurement depends on its priors, on the waveform, on the sampler hyperparameters, and on auxiliary inputs. We first validate the pipeline against GW150914 with the LVK GWTC-2.1+ production waveform \XPHM \citep{Pratten2021XPHM,Abbott2021GWTC2p1}. We then report the \IMRX\ baseline GW170817 posterior, compare it with a direct uniform-in-$d_L$ posterior and with the reweighted uniform-in-$d_L$ posterior, and use targeted mode-isolated runs to identify the $(\dL,\iota)$ bimodality that mechanistically explains the reweighting deficit. A waveform cross-check against \TF\ bounds the residual waveform systematic. Robustness sweeps over sampler hyperparameters, heterodyne-bin count, PSD source, heterodyne reference parameters, the host peculiar-velocity centre, and a companion \IMR\ prior sweep are referenced in Appendix~\ref{app:robustness}. We finish by reporting heterodyne speedup and live-point scaling.

Section~\ref{sec:method} describes the inference setup. Section~\ref{sec:validation} reports validation. Section~\ref{sec:results} contains the prior-sensitivity analysis, the bimodality characterisation, and the waveform cross-check. Section~\ref{sec:performance} reports runtime and live-point scaling. Section~\ref{sec:discussion} discusses implications and limitations. Section~\ref{sec:conclusions} concludes. Appendix~\ref{app:robustness} reports the inline robustness sweeps that are referenced in the body but not promoted to main figures.

\section{Method}
\label{sec:method}

\subsection{Nested sampling}

For data $d$ and source parameters $\theta$ under model $\mathcal{M}$, we target the posterior
\begin{equation}
p(\theta\mid d,\mathcal{M}) = \frac{\mathcal{L}(d\mid\theta,\mathcal{M})\,\pi(\theta\mid\mathcal{M})}{Z(d\mid\mathcal{M})},
\end{equation}
where $\mathcal{L}$ is the gravitational-wave likelihood, $\pi$ the prior, and $Z$ the Bayesian evidence. We estimate $Z$ and draw posterior samples by nested sampling \citep{Skilling2006Nested}, using the Nested Slice Sampling implementation of \citet{Yallup2026BlackjaxNS} in the \bjns\ framework \citep{BlackJAX}. The sampler evolves a population of live points through batched slice updates; this vectorisation keeps the GPU saturated when the per-evaluation likelihood cost is small.

All heterodyned GW170817 science runs reported here use $n_{\rm live}=5000$ live points, $n_{\rm delete}=2500$ points removed per iteration ($\tfrac12\,n_{\rm live}$), and $n_{\rm mcmc}=8\,n_{\rm dim}=112$ slice-sampling steps per update, where $n_{\rm dim}$ is the dimensionality of the sampled parameter space (Table~\ref{tab:sampler}); runs terminate when the fractional evidence increment is below $10^{-3}$. Scaling experiments vary $n_{\rm live}$ while holding all other settings fixed. The GW150914 \XPHM\ validation run uses the larger $n_{\rm live}=8000$, $n_{\rm delete}=2400$, $n_{\rm mcmc}=160$ ($16\,n_{\rm dim}$) configuration to verify that the validation conclusions are not sampler-limited at the smaller short-signal $n_{\rm dim}$. A dedicated $n_{\rm mcmc}$ convergence sweep at $\{5,10,20\}\times n_{\rm dim}$ for the 14-dimensional GW170817 problem is left to follow-up work, with the headline tail probability $P(\hzero>120\kmsmpc)$ as the natural convergence diagnostic.

\begin{table}
  \centering
  \caption{Nested sampling settings. $n_{\rm dim}$ is the sampled parameter-space dimension (14 for the phase-marginalised GW170817 runs, 10 for GW150914). Scaling experiments vary $n_{\rm live}$ with all other settings fixed.}
  \label{tab:sampler}
  \begin{tabular}{lcc}
    \toprule
    Setting & GW170817 & GW150914 \\
    \midrule
    $n_{\rm dim}$ (parameter dimension)      & $14$ & $10$ \\
    $n_{\rm live}$ (live points)             & $5000$ & $8000$ \\
    $n_{\rm delete}$ (removed per iteration) & $2500$ & $2400$ \\
    $n_{\rm mcmc}$ (slice steps per update)  & $112$ & $160$ \\
    \quad as a multiple of $n_{\rm dim}$     & $8\,n_{\rm dim}$ & $16\,n_{\rm dim}$ \\
    \bottomrule
  \end{tabular}
\end{table}

\subsection{Heterodyned likelihood}

The full frequency-domain likelihood for GW170817 uses $259\,201$ bins over $20$--$2048\,\rm Hz$ at $\Delta f=1/128\,\rm Hz$ for the $128\,\rm s$ strain segment. We use a heterodyned (relative-binning) likelihood \citep{Cornish2010,Zackay2018RelativeBinning}, in which the ratio between a candidate waveform and a reference waveform is approximated by a piecewise-linear function over a small set of coarse frequency bins; we use $501$ nominal bins for GW170817 and $383$ for GW150914. Other likelihood-compression schemes with different accuracy and setup trade-offs are available, notably reduced-order quadrature \citep{Smith2016ROQ} and multibanding or adaptive frequency resolution \citep{Vinciguerra2017Multiband,Morisaki2021Multiband}; we adopt relative binning for its simplicity and natural fit to a differentiable GPU likelihood. The reference waveform parameters and the per-waveform effective bin count after frequency-support masking are recorded in each run's \texttt{config.json}. We compare heterodyned and full-resolution posteriors directly in Section~\ref{sec:hetero-vs-unhet} to verify that the approximation does not bias the source-parameter posterior at the level needed for \hzero inference.

All heterodyned runs analytically marginalise over the coalescence phase, so the sampled parameter space is fourteen-dimensional for GW170817 (intrinsic masses, aligned-spin $z$-components, tidal deformabilities, inclination, luminosity distance, coalescence time, polarisation angle, sky position, and the cosmological parameters $\hzero$ and $v_{p}$), and ten-dimensional for GW150914.

\subsection{Joint Hubble-constant likelihood}

We sample the Hubble constant jointly with the gravitational-wave source parameters rather than post-processing the \dL\ marginal. The GW170817 host galaxy NGC~4993 contributes a CMB-frame recessional velocity $v_{\rm r}=3327\kms$ with $\sigma_{v_{\rm r}}=72\kms$ and a Hubble-flow peculiar-velocity estimate $\langle v_{p}\rangle=310\kms$ with $\sigmavp=150\kms$. Both enter as Gaussian likelihood factors,
\begin{equation}
\begin{split}
\ln\mathcal{L}_{\rm tot}(\theta,\hzero,v_{p}) = {}& \ln\mathcal{L}_{\rm GW}(d\mid\theta) \\
&+ \ln\mathcal{N}\!\left(v_{\rm r}\,;\,\hzero\,\dL+v_{p},\,\sigma_{v_{\rm r}}\right) \\
&+ \ln\mathcal{N}\!\left(\langle v_{p}\rangle\,;\,v_{p},\,\sigmavp\right) ,
\end{split}
\label{eq:h0likelihood}
\end{equation}
following \citet{Abbott2017H0}. Equation~\ref{eq:h0likelihood} carries no $1/N_{\rm s}(\hzero)$ GW-detection selection factor because, for every prior variant considered in this work, $\pi(\dL\mid\hzero)$ is independent of \hzero\ and the selection normalisation $N_{\rm s}=\int P_{\rm det}(\dL)\,\pi(\dL\mid\hzero)\,d\dL$ is therefore an \hzero-independent constant that cancels from the joint posterior; we return to this point with the prior definitions in Section~\ref{sec:priors}.

\subsection{Priors and prior variants}
\label{sec:priors}

The baseline prior configuration follows the bright-siren analysis of \citet{Abbott2017H0}: $\pi(\dL)\propto\dL^{2}$ over $[10,75]\,\rm Mpc$, $\pi(\hzero)\propto\hzero^{-1}$ over $[45,250]\kmsmpc$, and $\pi(v_{p})\propto 1$ over $[-1000,1000]\kms$ combined with the Gaussian $v_{p}$-likelihood term in equation~\ref{eq:h0likelihood} at $\sigmavp=150\kms$. The source-parameter priors reproduce those of the LVK GW170817 parameter estimation \citep{Abbott2019GW170817Properties}: the prior is uniform in the detector-frame component masses over $[0.5,7.7]\,M_{\odot}$ with $m_{1}\ge m_{2}$, subject to the additional chirp-mass constraint $1.184\le\mathcal{M}^{\rm det}\le 2.168\,M_{\odot}$ \citep[][Sec.~III\,D]{Abbott2019GW170817Properties}. As noted there, these bounds were chosen for technical reasons and the posterior has no support near them. The prior is full-sky, with right ascension uniform on $[0,2\pi]$ and declination distributed according to a $\cos\delta$ prior on $[-\pi/2,\pi/2]$. Every heterodyned run reported here is full-sky.

We consider three prior variants in addition to the baseline:
\begin{enumerate}
    \item \emph{Direct uniform-in-$d_L$}: $\pi(\dL)$ uniform over the same fixed $[10,75]\,\rm Mpc$ window (independent of \hzero), sampled directly.
    \item \emph{Reweighted uniform-in-$d_L$}: the baseline posterior samples reweighted by the analytic ratio of the uniform-in-$d_L$ to volumetric \dL\ priors.
    \item \emph{Enlarged peculiar-velocity uncertainty}: identical to the baseline except $\sigmavp=250\kms$.
\end{enumerate}
The direct and reweighted uniform-in-$d_L$ posteriors share the same target prior; they differ only in whether the prior is imposed during sampling or applied post-hoc. At GW170817's source redshift ($z\lesssim 0.02$), the uniform-in-$d_L$ prior is numerically equivalent within one per cent to the ``flat-in-redshift'' variant analysed by \citet{Abbott2017H0} over the same distance range; we adopt the literal uniform-in-$d_L$ form because it is what the inference pipeline imposes ($\pi(\dL)$ is a fixed density on the observable \dL, with no \hzero-dependent window). Consequently $\pi(\dL\mid\hzero)$ has no \hzero-dependence for any of the three prior variants, so the GW-detection selection normalisation $N_{\rm s}(\hzero)$ of equation~\ref{eq:h0likelihood} is constant in \hzero\ and cancels from the joint posterior, exactly as for the volumetric baseline.\footnote{\raggedright We verified this property by direct evaluation of $N_{\rm s}(\hzero)=\int P_{\rm det}(\dL)\,\pi(\dL\mid\hzero)\,d\dL$ over the LVK-matched $\hzero\in[45,250]\kmsmpc$ prior with a Finn--Chernoff antenna-pattern detection model \citep{FinnChernoff1993}: $N_{\rm s}$ is \hzero-independent to machine precision for every horizon distance tested ($D_{\rm h}\in\{100,150,220\}\,\rm Mpc$, spanning the published O2 BNS range to inspiral horizon; \citealp{LVK2020Prospects}). The supporting calculation is in \texttt{analyze\_selection\_term.py} in the public repository.}

\subsection{Waveforms}
\label{sec:waveforms}

For GW170817 we adopt two waveforms in the main analysis, both evaluated with the \jax\ \ripple\ implementation \citep{Edwards2023Ripple} within the \textsc{jim} framework \citep{Wong2023Jim}. The primary is \IMRX, which combines the most recent NR-calibrated tidal-phase prescription \citep[NRTidalv3;][]{Abac2024NRTidalv3} with a modern aligned-spin base \citep[IMRPhenomXAS;][]{Pratten2020XAS}; the differentiable \jax\ implementation we use is introduced in \citet{Chan2026ripple}.\footnote{\raggedright The \IMRX\ waveform is evaluated with R.~Chan's \jax\ reimplementation of the IMRPhenomX family in \ripple\ (\url{https://github.com/GW-JAX-Team/ripple}), used through the \textsc{jim} framework \citep{Wong2023Jim} (\url{https://github.com/GW-JAX-Team/jim}); its fidelity against the \textsc{LALSuite} reference implementation will be characterised in \citet{Chan2026ripple}.} \TF\ is included as a post-Newtonian inspiral-only family check, and we additionally analysed \IMR\ as an anchor waveform for the IMRX-vs-NRTidalv2 capture-fraction comparison reported inline in Section~\ref{sec:prior}; the \jax\ implementations of both tidal models were introduced in \citet{Wouters2024JimBNS}, and the corresponding full \IMR\ posteriors are retained in the public test suite and summarised in Appendix~\ref{app:robustness}.

For GW150914 validation we use \XPHM \citep{Pratten2021XPHM}, the LVK GWTC-2.1+ production waveform for that event, which incorporates precession through a multi-scale-analysis treatment together with the dominant higher-order modes; we evaluate it with the \jax\ \ripple\ implementation \citep{Edwards2023Ripple} within the \textsc{jim} framework \citep{Wong2023Jim}. Like-for-like comparison against the public LVK PE \citep{Abbott2021GWTC2p1,GWTC2p1_GW150914_Zenodo} therefore requires no waveform substitution.

\subsection{Posterior summaries}
\label{sec:summaries}

We summarise one-dimensional marginals by a point estimate and a credible interval, both computed directly from the weighted posterior samples. The point estimate is the weighted-histogram mode, which we denote the maximum a posteriori (MAP) value: the centre of the maximum-weighted-count bin on a fixed $1\kmsmpc$ grid over $[40,230]\kmsmpc$. This is a posterior mode estimated from the sample set, not a likelihood maximisation; for the skewed uniform-in-$d_L$ variants in Table~\ref{tab:h0priors} we additionally report the weighted posterior median (the interpolated $50$th percentile of the weight-ordered cumulative distribution) as a bin-noise-robust complement. The credible intervals are highest-posterior-density (HPD) intervals -- the shortest contiguous interval containing the requested cumulative weight -- computed directly from the weighted samples at the stated credibility. All one-dimensional \emph{summaries} -- the MAP, the weighted median, the HPD intervals, and the tail probabilities $P(\hzero>120\kmsmpc)$ that carry the prior-sensitivity signal of Section~\ref{sec:prior} -- are computed directly from the weighted samples with no kernel-density smoothing, so the reported tail statistics are not a smoothing artefact. The one-dimensional marginal \emph{curves} displayed in Figures~\ref{fig:h0prior}, \ref{fig:bimodality} and \ref{fig:waveform-h0} use a Silverman-bandwidth kernel-density estimate for visual presentation only; the quantitative tail comparisons never rely on them. Two-dimensional corner contours are likewise KDE-smoothed.

\section{Validation}
\label{sec:validation}

\subsection{GW150914}
\label{sec:gw150914}

\begin{figure}
  \centering
  \includegraphics[width=\columnwidth]{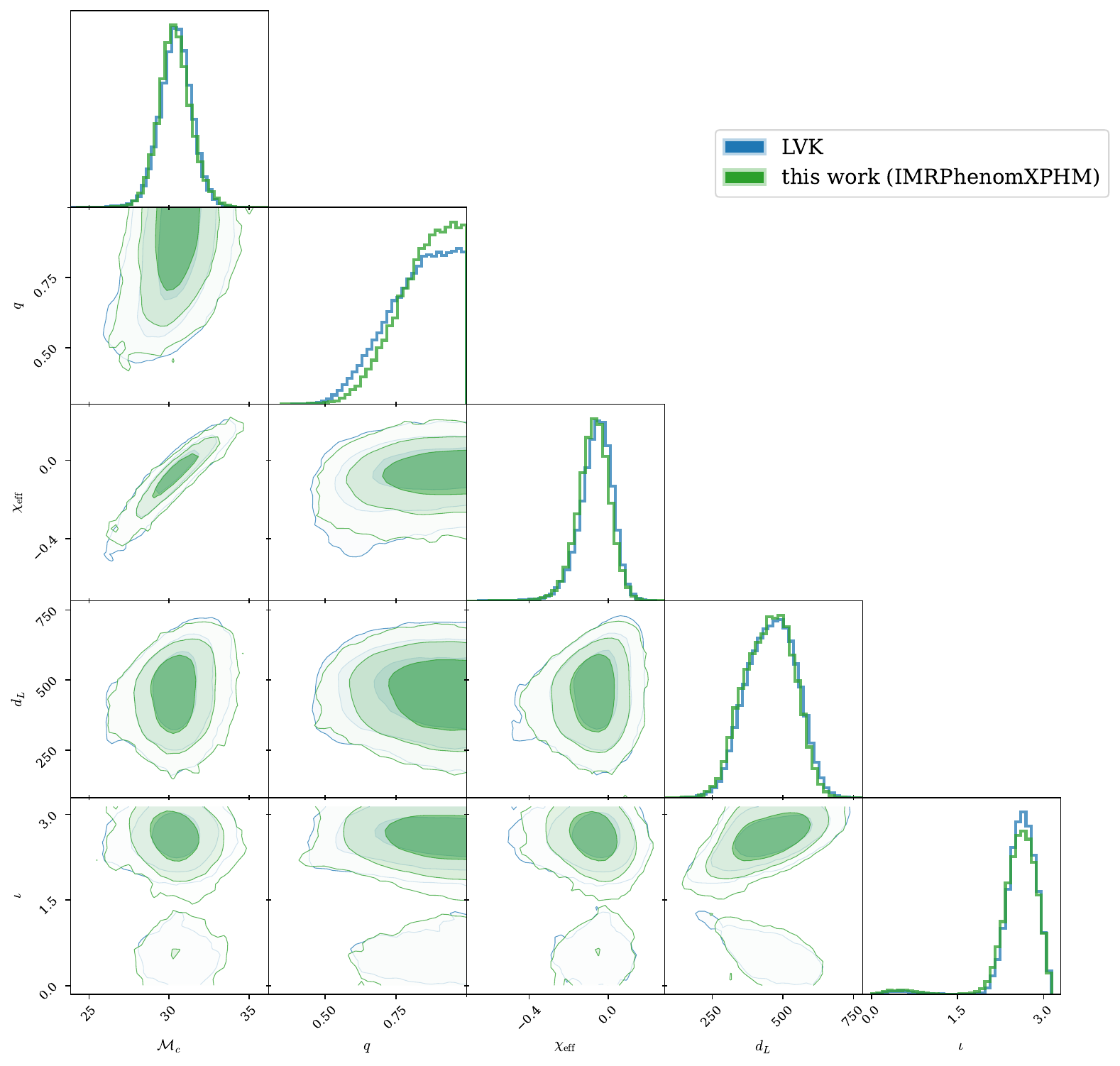}
  \caption{GW150914 validation: corner overlay of the LVK GWTC-2.1 \XPHM\ posterior \citep{Abbott2021GWTC2p1,GWTC2p1_GW150914_Zenodo} and our heterodyned \XPHM\ run at $n_{\rm live}=8000$, $n_{\rm mcmc}=160$. The two posteriors overlap throughout the recovered support; our component-mass prior $[5,100]\,M_\odot$ encompasses the LVK $[10,80]\,M_\odot$ range without adding posterior mass at the boundaries.}
  \label{fig:gw150914}
\end{figure}

We validate the pipeline on the binary-black-hole event GW150914 \citep{Abbott2016GW150914}, which has a short signal and well-established public posteriors. We analyse the public GW150914 strain data and PSDs from the LVK GWTC-2.1 Zenodo deposit \citep{GWTC2p1_GW150914_Zenodo} with \XPHM, using the same heterodyned sampler as for GW170817 with phase marginalisation;\footnote{\raggedright Phase marginalisation is not formally valid for a precessing, higher-multipole waveform such as \XPHM, because higher-order modes break the single overall-phase dependence that analytic phase marginalisation assumes. We retain it here for consistency with the GW170817 configuration; the agreement with the published LVK PE -- which does not marginalise over phase -- throughout the recovered support (Table~\ref{tab:gw150914}, Figure~\ref{fig:gw150914}) confirms it does not measurably bias this validation. The GW170817 science runs use aligned-spin tidal waveforms, for which phase marginalisation is formally valid, so this choice does not affect the \hzero\ posteriors.} for this validation run we increase the sampler settings to $n_{\rm live}=8000$ with $n_{\rm mcmc}=160$ slice steps per update, so that the comparison is not sampler-limited at the short-signal $n_{\rm dim}$. Table~\ref{tab:gw150914} compares our recovered parameters with the LVK GWTC-2.1 public PE \citep{Abbott2021GWTC2p1,GWTC2p1_GW150914_Zenodo}; the corresponding corner overlay is shown in Figure~\ref{fig:gw150914}. The chirp mass, mass ratio, luminosity distance, and inclination all agree with the LVK reference; in particular, the recovered chirp mass agrees to ${\sim}0.2\,\%$ ($\mathcal{M}_c=30.35$ vs $30.4\,M_\odot$) and $d_{L}^{\rm med}=455\,\rm Mpc$ for our \XPHM\ run is within ${\sim}8\,\rm Mpc$ of the LVK value of $463\,\rm Mpc$. The 1$\sigma$ widths match closely: $\sigma(\mathcal{M}_c)=1.02$ vs $1.00\,M_\odot$, $\sigma(d_L)=84$ vs $85\,\rm Mpc$, $\sigma(q)=0.10$ vs $0.11$, and $\sigma(\iota)=0.41$ vs $0.35\,\rm rad$ (ratios $1.02$, $0.99$, $0.89$, $1.17$); the largest residual is on the precession-sensitive $\iota$. We use a component-mass prior uniform on $[5,100]\,M_\odot$ that encompasses the LVK GWTC-2.1 $[10,80]\,M_\odot$ range, with no posterior mass near either boundary; the validation therefore reproduces the LVK posteriors within the recovered support.

\begin{table*}
  \centering
  \caption{GW150914 validation: median values of the principal source parameters for our heterodyned \XPHM\ run, with a $n_{\rm live}=5000$ cross-check. The LVK GWTC-2.1 \XPHM\ public PE \citep{Abbott2021GWTC2p1,GWTC2p1_GW150914_Zenodo} reports $\mathcal{M}_c\approx 30.4\,M_\odot$, $q\approx 0.85$, $d_L\approx 463\,\rm Mpc$, $\iota\approx 2.62\,\rm rad$ for direct comparison.}
  \label{tab:gw150914}
  \begin{tabular}{lccccc}
    \toprule
    Run & $\mathcal{M}_{c}/M_{\odot}$ & $q$ & $d_{L}/\rm Mpc$ & $\iota/\rm rad$ & $\ln Z$ \\
    \midrule
    \XPHM\ ($n_{\rm live}=8000$)          & 30.35 & 0.87 & 455 & 2.61 & $260.86\pm0.06$ \\
    \XPHM\ ($n_{\rm live}=5000$, cross-check) & 30.34 & 0.87 & 460 & 2.62 & $261.09\pm0.08$ \\
    \bottomrule

  \end{tabular}
\end{table*}

\subsection{Consistency of the likelihood compression on GW170817}
\label{sec:hetero-vs-unhet}

\begin{figure}
  \centering
  \includegraphics[width=\columnwidth]{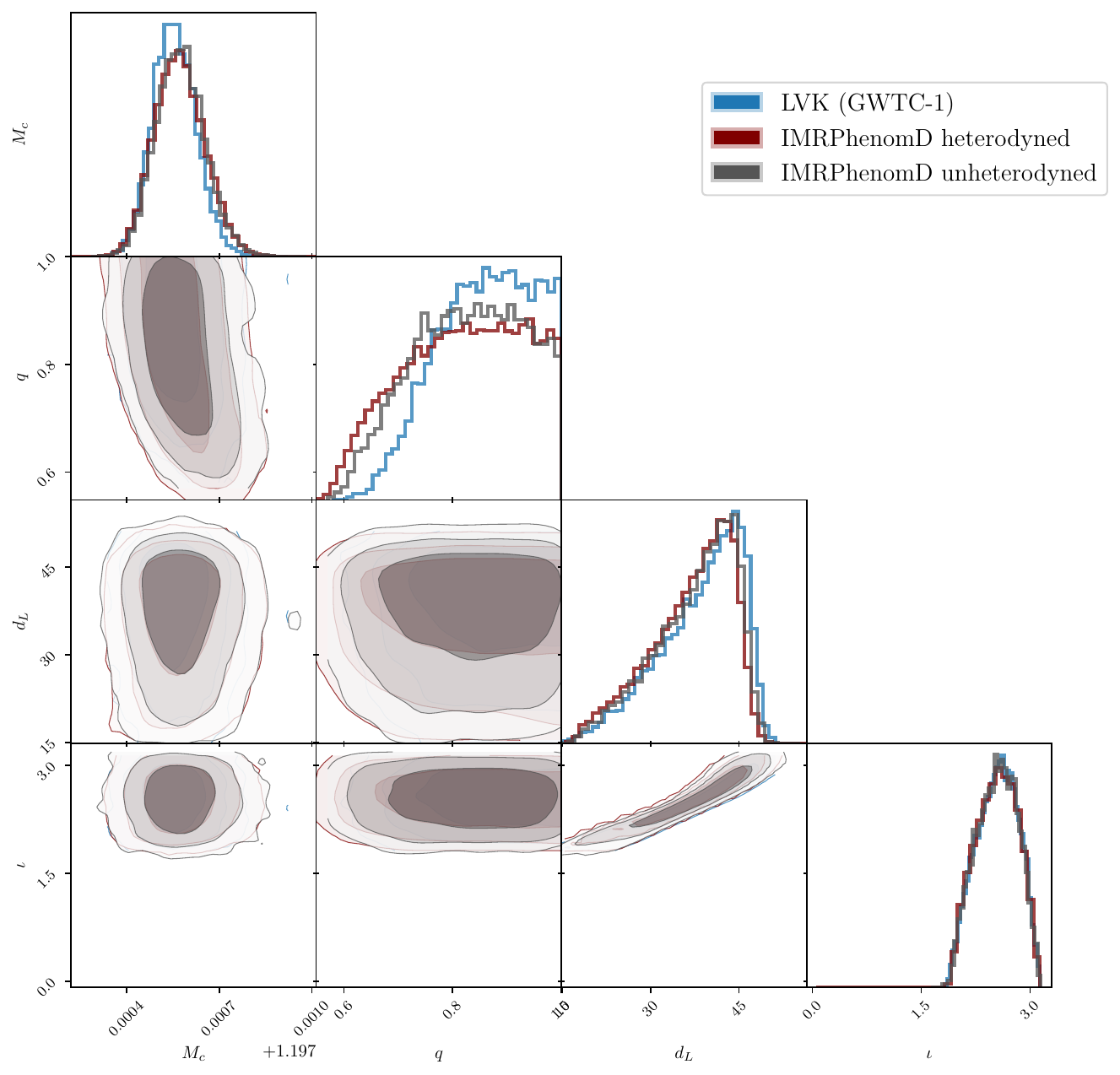}
  \caption{Heterodyned vs full-resolution consistency for GW170817 under \IMR\ at $n_{\rm live}=1500$. Source-parameter corner overlay of the heterodyned (relative-binning) and full-resolution (full $259\,201$-bin) posteriors, with the public GWTC-1 IMRPhenomPv2\_NRTidal posterior \citep{Abbott2019GWTC1,LVK_GW170817_DataRelease} shown for reference. The heterodyned and full-resolution $(\mathcal{M}_{c},q,\dL,\iota)$ contours overlap throughout, confirming that the relative-binning approximation does not bias the source-parameter posterior at the level needed for \hzero\ inference. Two-dimensional contours and one-dimensional marginals are KDE-smoothed.}
  \label{fig:hetero-vs-unhet}
\end{figure}

We next verify that the heterodyned (relative-binning) approximation does not bias the GW170817 source-parameter posterior. Relative binning is by now an established compression technique for long-duration binary-neutron-star signals -- the DINGO-BNS framework, for example, adopts heterodyning to compress exactly this class of signal \citep{Dax2025DingoBNS}, and rapid approximate analyses are routine for low-latency alerts -- but we confirm it directly for our configuration. We run the same sampler on the full-resolution $259\,201$-bin frequency-domain likelihood for both \IMR\ and \TF\ at $n_{\rm live}=1500$, and additionally at $n_{\rm live}\in\{500,2500\}$ for \IMR. The heterodyned and full-resolution $\dL$, $\iota$, and $\hzero$ marginals overlay closely (Figure~\ref{fig:hetero-vs-unhet}), and the heterodyned $\hzero^{\rm MAP}$ lies between the surrounding full-resolution cross-checks at the level of the sampler-statistical width. We conclude that the heterodyned approximation is adequate for \hzero inference at this event's signal-to-noise ratio; the corresponding wall-clock reduction is quantified in Section~\ref{sec:speedup}.

\section{Results}
\label{sec:results}

\subsection{Prior sensitivity of the GW170817 \hzero\ posterior}
\label{sec:prior}

\begin{figure}
  \centering
  \includegraphics[width=\columnwidth]{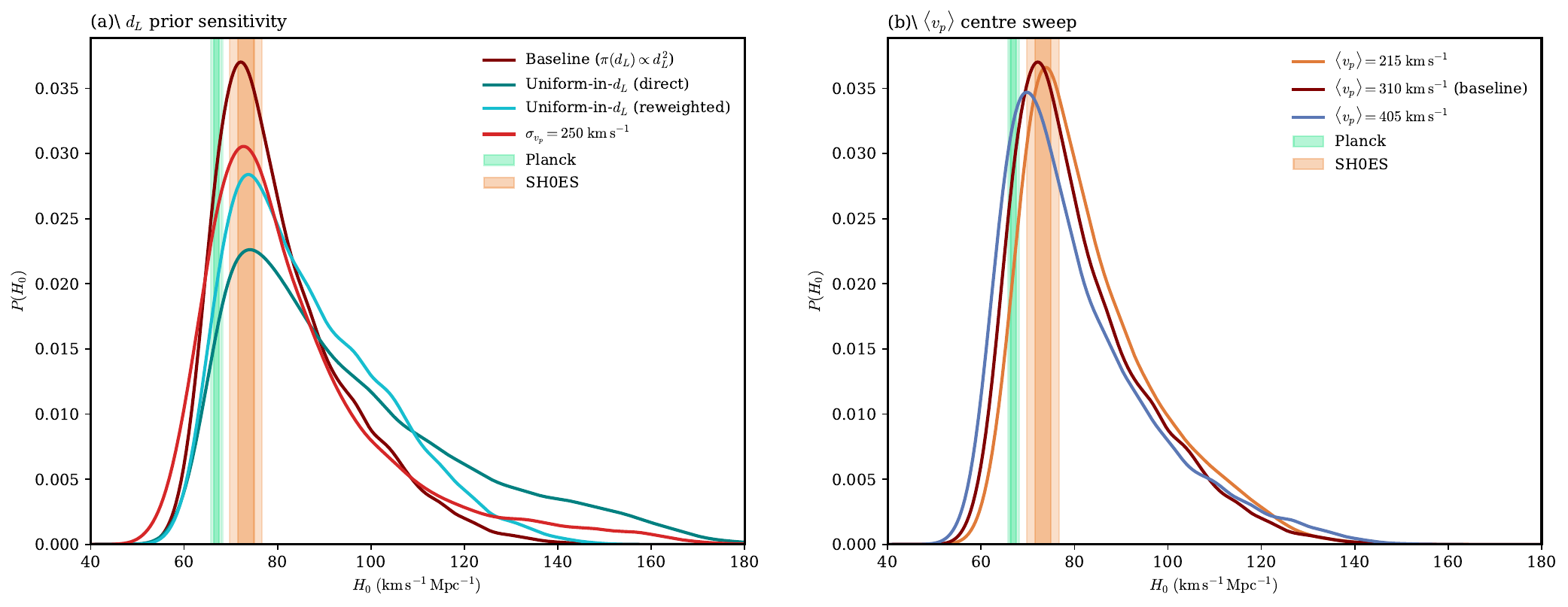}
  \caption{Prior-sensitivity comparison for \IMRX. \emph{Left panel (a):} kernel-density estimates of the \hzero\ posterior under the four distance-prior variants of Section~\ref{sec:priors} (volumetric baseline, direct uniform-in-$d_L$, reweighted uniform-in-$d_L$, and $\sigmavp=250\kms$). The directly sampled uniform-in-$d_L$ posterior places substantially more mass at high \hzero\ than the reweighted estimate, despite both targeting the same prior. \emph{Right panel (b):} peculiar-velocity centre sweep --- \hzero\ posteriors for $\langle v_{p}\rangle\in\{215,310,405\}\kms$ at fixed $\sigmavp=150\kms$. Planck $\Lambda$CDM and SH0ES distance ladder bands are shown in both panels.}
  \label{fig:h0prior}
\end{figure}

The central scientific result of this paper is a direct comparison between the uniform-in-$d_L$ \hzero\ posterior obtained by sampling under the target prior and that obtained by reweighting the baseline posterior to the same target prior, under the locked primary waveform \IMRX. We carry out the four prior variants of Section~\ref{sec:priors} on the LVK-matched full-sky prior set and report sample-derived HPDs in Table~\ref{tab:h0priors}; the corresponding marginals are shown in Figure~\ref{fig:h0prior}.

\begin{table*}
  \centering
  \caption{\IMRX\ \hzero\ summaries on the LVK-matched full-sky prior set, $n_{\rm live}=5000$. Upper block: the four distance-prior variants of Section~\ref{sec:priors} (volumetric baseline, direct and reweighted uniform-in-$d_L$, and $\sigmavp=250\kms$). Lower block: the peculiar-velocity centre sweep $\langle v_p\rangle\in\{215,405\}\kms$ at fixed $\sigmavp=150\kms$ (the $\langle v_p\rangle=310\kms$ point is the baseline). MAP, weighted median, and sample-derived 68 per cent HPDs in $\kmsmpc$; the median is reported as a bin-noise-robust complement to the MAP for the skewed posteriors of the uniform-in-$d_L$ variants. The reweighted variant is post-hoc and has no independent $\ln Z$. All numbers generated by \texttt{Plots/build\_paper\_tables.py} from the canonical sample CSVs.}
  \label{tab:h0priors}
  \begin{tabular}{lcccccc}
    \toprule
    & \multicolumn{3}{c}{$\hzero~(\kmsmpc)$} & & & \\
    \cmidrule(lr){2-4}
    Variant & MAP & median & 68\% HPD & $P(\hzero>120)$ & $P(\hzero>150)$ & $\ln Z$ \\
    \midrule
    Baseline ($\pi(d_L)\propto d_L^{2}$)           & 70.5 & 77.6 & $[63.8,87.6]$ & $0.017$ & $<10^{-4}$ & $486.25\pm0.11$ \\
    Uniform-in-$d_L$, direct                       & 70.5 & 87.6 & $[64.2,103.8]$ & $0.159$ & $0.038$ & $487.30\pm0.10$ \\
    Uniform-in-$d_L$, reweighted                   & 73.5 & 82.9 & $[65.2,95.9]$ & $0.041$ & $0.000$ & (post-hoc) \\
    $\sigma_{v_p}=250\,\mathrm{km\,s^{-1}}$        & 73.5 & 78.3 & $[61.7,90.5]$ & $0.069$ & $0.015$ & $485.55\pm0.09$ \\
    \midrule
    $\langle v_p\rangle=215\,\mathrm{km\,s^{-1}}$  & 74.5 & 79.9 & $[66.1,90.4]$ & $0.021$ & $<10^{-4}$ & $485.76\pm0.12$ \\
    $\langle v_p\rangle=405\,\mathrm{km\,s^{-1}}$  & 68.5 & 76.3 & $[61.5,87.0]$ & $0.030$ & $0.000$ & $486.35\pm0.10$ \\
    \bottomrule

  \end{tabular}
\end{table*}

\subsubsection{Direct-vs-reweighted comparison} The direct uniform-in-$d_L$ run leaves the posterior MAP unchanged at $70.5\kmsmpc$ but broadens the 68 per cent HPD on the high side from $87.6$ to $103.8\kmsmpc$, and raises the high-tail probability
\begin{equation*}
P(\hzero>120\kmsmpc) =
\begin{cases}
0.017 & \text{(baseline)}, \\
0.159 & \text{(direct uniform-in-}d_L\text{)} .
\end{cases}
\end{equation*}
together with $P(\hzero>150\kmsmpc)$ from $<10^{-4}$ to $0.038$. The reweighted estimate captures a small fraction of this shift: $P(\hzero>120\kmsmpc)=0.041$, equivalent to
\begin{equation}
\frac{0.041 - 0.017}{0.159 - 0.017} \approx 17\,\%
\label{eq:reweight-fraction}
\end{equation}
of the direct prior-induced change. The 68 per cent HPD upper bound under reweighting reaches $95.9\kmsmpc$, short of the directly sampled $103.8\kmsmpc$ at the same credibility. The reweighted variant is itself consistent with the flat-in-redshift posterior reported by \citet{Abbott2017H0} (${\approx}$ uniform-in-$d_L$ at this $z$; their Extended Data Table~1 gives a $68.3\,\%$ HPD $\hzero=71_{-9}^{+23}\kmsmpc$) -- so our reweighting reproduces the original analysis; the deficit we report lies in the directly sampled high-\hzero\ tail, which the reweighted procedure does not access. The three \IMRX\ prior variants that carry an independent evidence agree to $\Delta\ln Z\lesssim 1.8$; the decision-relevant pair, baseline versus direct uniform-in-$d_L$, differs by only $\Delta\ln Z\approx 1.05$. On the Jeffreys scale this is at most weak-to-substantial support, and it is within the run-to-run variability seen in the public repository for this configuration. The data therefore show at most a weak preference among the distance priors: the change in the high-\hzero\ tail across these variants is dominated by the \emph{prior}, not by a data-driven update.

\subsubsection{Reweighting bias versus variance} The reweighting deficit is bias, not high variance. A non-parametric bootstrap on the reweighted estimator (4\,000 multinomial-at-$n_{\rm eff}$ draws at the reweighted $n_{\rm eff}=27{,}317$) gives a $95$ per cent confidence interval for $P(\hzero>120\kmsmpc)$ of $[0.037, 0.042]$ -- tight to three significant figures and excluding the directly sampled $0.159$ by ${\sim}\,100$ binomial standard errors of the reweighted estimator. The conclusion is robust to the choice of resampler: a Bayesian (Dirichlet-weighted) bootstrap and a $200$-block jackknife (the standard nonparametric variance estimator for importance-sampling estimators; \citealp{Owen2013MonteCarlo}) both give intervals about $30\,\%$ wider but with the same centre, and still exclude $0.159$ by more than $70$ nonparametric standard errors. Down-sampling the baseline to $S\in\{10^{4},3\times 10^{4},10^{5}\}$ draws leaves the reweighted point estimate within $0.005$ of $0.04$ and the bootstrap interval shrinking around $0.04$ rather than around $0.159$: the reweighted estimator is converging on the wrong limit, not converging slowly. Only direct sampling under the target prior recovers the high-\hzero\ tail mass. The bootstrap, the sample-size sweep, and the corresponding $\hat{k}$ value below are computed by \texttt{analyze\_psis\_khat.py} in the public repository and recorded in the \texttt{paper\_diagnostics.csv} summary.

\subsubsection{Effective-sample-size diagnostic} The reweighting failure is independently flagged by the effective sample size of the reweighted posterior. We use Kish's effective sample size, $n_{\rm eff}=\left(\sum_i w_i\right)^2/\sum_i w_i^2$ \citep{Kish1965}, where $w_i$ are the (nested sampling or importance) weights. From the same $1.73\times 10^{5}$-sample \IMRX\ baseline we measure $n_{\rm eff}=30{,}695$ (efficiency $17.8\,\%$); the directly sampled uniform-in-$d_L$ run has $n_{\rm eff}=37{,}022$ from its own $1.94\times 10^{5}$-sample draw (efficiency $19.1\,\%$), while the reweighted uniform-in-$d_L$ estimator has $n_{\rm eff}=27{,}317$ -- \emph{lower} than the baseline despite reweighting from the same draw. The reweighted-versus-baseline comparison is the coverage diagnostic; the directly sampled run is listed for context only, as it has a different sampling history. This is the conventional symptom of importance-sampling weights concentrated on a small subset of the baseline draw, the regime in which reweighting is known to be unreliable \citep{Payne2019Reweighting,Vehtari2024PSIS}. The Pareto-smoothed importance-sampling $\hat{k}$ diagnostic of \citet{Vehtari2024PSIS}, computed by a generalised-Pareto maximum-likelihood fit to the top $M=3\sqrt{S}=1{,}264$ log-importance ratios (the default cap of \citealp{Vehtari2024PSIS}), gives $\hat{k}=0.683$ for the reweighted \IMRX\ draw: at the upper edge of the cautionary $0.5<\hat{k}\le 0.7$ ``high variance but consistent'' regime, just below the standard $\hat{k}>0.7$ unreliability threshold.\footnote{\raggedright We cross-checked the local GPD-MLE against the canonical \texttt{arviz.psislw} implementation under its PSIS-LOO sign convention, and the two agree to $10^{-3}$ ($\hat{k}_{\rm arviz}=0.683$); the value is also insensitive to the tail-fraction choice within the $M=3\sqrt{S}$ Vehtari cap.} The empirical bootstrap above shows the bias is already severe at this borderline $\hat{k}$, so for prior-sensitivity reweighting in the broad-posterior bright-siren regime we recommend reporting a bootstrap confidence interval alongside $\hat{k}$ and $n_{\rm eff}$. All three diagnostics are cheap, require no additional GPU time, and would have flagged the deficit in the original analysis from the baseline draw alone.

\subsubsection{Peculiar-velocity sweep} The $\sigmavp=250\kms$ variant shifts the \hzero\ posterior modestly: $P(\hzero>120)$ moves from $0.017$ to $0.069$. Sweeping the $v_{p}$ prior \emph{centre} over the historical literature range $\langle v_{p}\rangle\in\{215, 310, 405\}\kms$ at fixed $\sigmavp=150\kms$ (Figure~\ref{fig:h0prior}b, Table~\ref{tab:h0priors}) shifts $\hzero^{\rm MAP}$ by $6\kmsmpc$ peak-to-peak ($74.5$ at $\langle v_{p}\rangle=215\kms$ to $68.5$ at $\langle v_{p}\rangle=405\kms$) while $P(\hzero>120)$ changes by less than $0.02$. The $\langle v_{p}\rangle=310\kms$ run replicates the \IMRX\ baseline to within $0.01$ in $\ln Z$. For GW170817 with NGC~4993 identified, the dominant prior-induced uncertainty in \hzero\ is therefore the distance prior, not the peculiar-velocity prior.

\subsubsection{Cross-waveform check} Repeating the four-variant prior sweep on the legacy aligned-spin tidal model \IMR\ (NRTidalv2 calibration) gives qualitatively the same behaviour with a larger amplitude: baseline $P(\hzero>120)=0.076$; uniform-in-$d_L$ direct $P=0.281$; reweighted $P=0.195$; reweighting capture fraction $(0.195-0.076)/(0.281-0.076)\approx 58\,\%$. The newer NRTidalv3 calibration in \IMRX\ tightens the upper \hzero\ tail relative to NRTidalv2, leaving less Mode-B mass for the prior shift to redistribute and therefore both reducing the absolute prior-induced change in $P(\hzero>120)$ \emph{and} reducing the fraction of that change that reweighting can recover. The qualitative conclusion -- that reweighting systematically understates the prior-induced change in the high-\hzero\ tail -- is therefore robust across the tidal-calibration axis, and is in fact \emph{more} severe under the modern waveform than under the older one.

The shifts reported here are methodological, not cosmological: GW170817 alone is consistent with both early- and late-Universe \hzero\ measurements under every prior considered. The point is that the magnitude of this prior-induced change is a property of the inference, and reweighting alone reports a materially smaller value than direct sampling does.

\subsection{The $d_{L}$--$\iota$ bimodality}
\label{sec:bimodality}

\begin{figure}
  \centering
  \includegraphics[width=\columnwidth]{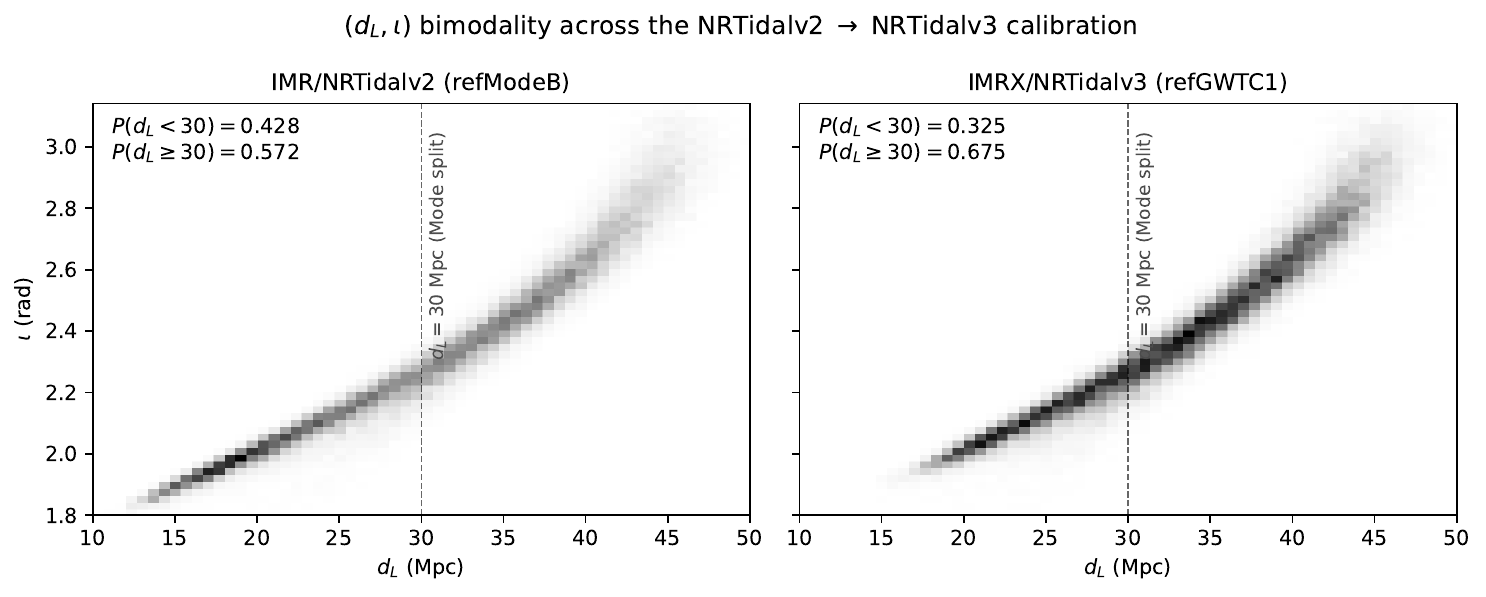}
  \caption{Cross-waveform check on the $(\dL,\iota)$ bimodality: weighted joint posteriors from the unrestricted direct uniform-in-$d_L$ runs at IMRPhenomD\_NRTidalv2 (left; Mode-B-anchored heterodyne reference) and \IMRX\ (right; default GWTC-1 reference). Both show the same two-peak structure with the Mode-B region $\dL<30\,\rm Mpc$ carrying $0.428$ (\IMR) and $0.325$ (\IMRX) of the posterior weight: the bimodality is robust across the NRTidalv2 $\to$ NRTidalv3 calibration. The smaller IMRX Mode-B weight is consistent with the tighter upper \hzero\ tail under NRTidalv3 reported in Section~\ref{sec:prior}.}
  \label{fig:bimodality-waveform-check}
\end{figure}

The uniform-in-$d_L$ posterior is broad and asymmetric, with non-negligible weight at $\dL\lesssim 25\,\rm Mpc$. Before isolating the modes quantitatively, we note that the same $(\dL,\iota)$ two-peak structure is visible in \emph{both} the IMRX and the \IMR\ direct uniform-in-$d_L$ posteriors (Figure~\ref{fig:bimodality-waveform-check}): the posterior weight in the Mode-B region $\dL<30\,\rm Mpc$ is $0.325$ under \IMRX\ and $0.428$ under \IMR. The bimodality is therefore a property of the data and the uniform-in-$d_L$ prior, not the NRTidalv2 tidal-phase calibration; the reduction in Mode-B weight from \IMR\ to \IMRX\ is consistent with the tighter upper \hzero\ tail under NRTidalv3 reported in Section~\ref{sec:prior}. The mode-isolated runs that follow use \IMR/NRTidalv2 at $n_{\rm live}=5000$; the cross-waveform check above shows the bimodality is robust to the tidal calibration, so the mode-isolated Bayes factor on \IMR\ characterises the same structure that drives the \IMRX\ reweighting deficit reported in Section~\ref{sec:prior}.

\begin{figure*}
  \centering
  \includegraphics[width=0.92\textwidth]{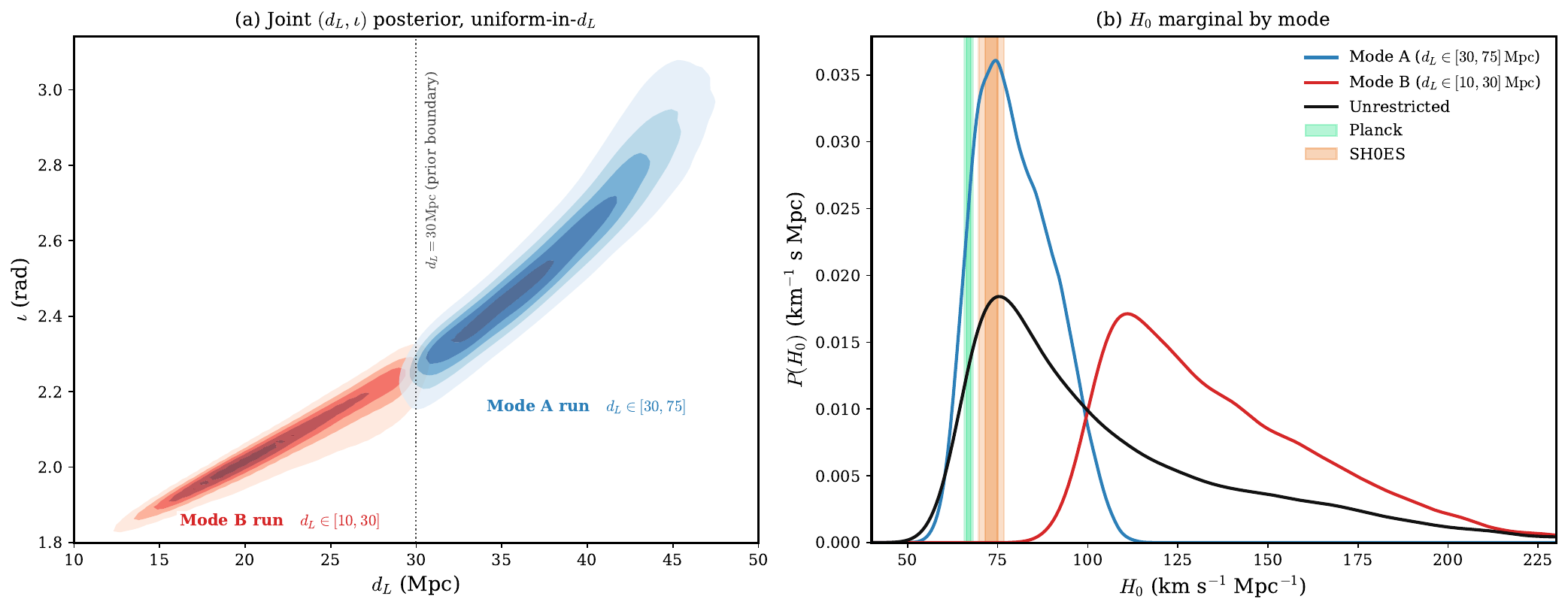}
  \caption{The $\dL$--$\iota$ bimodality in the GW170817 uniform-in-$d_L$ posterior. Left: joint $(\dL,\iota)$ posterior from the unrestricted $\dL\in[10,75]\,\rm Mpc$ run (Mode-B-anchored heterodyne reference); the prior-restricted Mode-A and Mode-B contours from the two separate restricted-prior runs are overlaid in distinct colours, with the $\dL=30\,\rm Mpc$ prior boundary marked. Right: per-mode \hzero\ marginals as Silverman-bandwidth kernel-density estimates. The Planck $\Lambda$CDM and SH0ES distance ladder bands are shown for context.}
  \label{fig:bimodality}
\end{figure*}

To resolve whether the \IMR\ mode structure is a single broadened mode or a genuine bimodality, we run two prior-restricted variants of the direct uniform-in-$d_L$ analysis: a Mode-A run with $\dL\in[30,75]\,\rm Mpc$ and a Mode-B run with $\dL\in[10,30]\,\rm Mpc$. The $\dL=30\,\rm Mpc$ split is placed at the approximate saddle separating the two maxima in the joint $(\dL,\iota)$ posterior (Figure~\ref{fig:bimodality}). We retain the original GWTC-1 heterodyne reference, and additionally run an unrestricted $\dL\in[10,75]\,\rm Mpc$ uniform-in-$d_L$ analysis with the heterodyne reference re-anchored at Mode~B ($\dL=20\,\rm Mpc$, $\iota=2.0\,\rm rad$) to test whether the apparent Mode-B mass is a heterodyne-reference artefact.

Figure~\ref{fig:bimodality} shows the joint $(\dL,\iota)$ posterior under the unrestricted run, together with the per-mode \hzero\ marginals. The support is a curved arm running from $(\dL,\iota)\approx(15,1.9)\,\rm Mpc{\cdot}rad$ to $(40,2.7)\,\rm Mpc{\cdot}rad$ with two distinct local maxima. Table~\ref{tab:bimodality} reports the local evidences and per-mode \hzero\ summaries.

\begin{table*}
  \centering
  \caption{Local evidences, MAP $\hzero$ values, and sample-derived 68 per cent HPDs for the $\dL$-restricted uniform-in-$d_L$ runs; the lower block is an independent replication. The Mode-B/Mode-A Bayes factor includes a $\ln(20/45)$ correction for the difference in restricted-prior $\dL$ volumes.}
  \label{tab:bimodality}
  \begin{tabular}{lccccc}
    \toprule
    Variant & $\dL$ range (Mpc) & $\hzero^{\rm MAP}\,(\kmsmpc)$ & 68\% HPD & $P(\hzero>120)$ & $\ln Z$ \\
    \midrule
    Mode A         & $[30,75]$ & 74.5 & $[66.7,88.8]$ & $<10^{-4}$ & $486.80\pm0.10$ \\
    Mode B         & $[10,30]$ & 109.5 & $[99.5,151.8]$ & $0.638$ & $486.95\pm0.09$ \\
    Unrestricted   & $[10,75]$ & 73.5 & $[62.6,116.5]$ & $0.281$ & $486.48\pm0.09$ \\
    \midrule
    Mode A         & $[30,75]$ & 72.5 & $[66.1,88.6]$ & $<10^{-4}$ & $486.71\pm0.09$ \\
    Mode B         & $[10,30]$ & 110.5 & $[99.0,154.5]$ & $0.646$ & $487.62\pm0.10$ \\
    Unrestricted   & $[10,75]$ & 73.5 & $[62.0,121.1]$ & $0.311$ & $487.52\pm0.09$ \\
    \bottomrule

  \end{tabular}
\end{table*}

The Bayes factor between Mode~B and Mode~A, after correcting for the volume difference between the restricted priors, is
\begin{equation*}
\begin{split}
\ln \mathcal{B}_{\rm B/A} &= (\ln Z_{\rm B}-\ln Z_{\rm A}) + \ln(20/45) \\
&= +0.15 - 0.81 = -0.66 \quad (\text{seed=0}).
\end{split}
\end{equation*}
An independent second-seed replication (Table~\ref{tab:bimodality}, lower block) gives $+0.91 - 0.81 = +0.10$ (seed=1). The $\ln(20/45)$ term is the log-ratio of the restricted-prior $\dL$ widths; it is the exact volume correction because the prior is uniform in $\dL$ by construction. Both seeds satisfy $|\ln\mathcal{B}_{\rm B/A}|<1$, placing the result in the ``not worth more than a mention'' range of the Jeffreys scale. We caution that the run-to-run $\ln Z$ scatter is larger than the nominal $\pm 0.1$ per-run uncertainty: the unrestricted $\ln Z$ differs by $1.04$ between the two seeds, so the $\pm 0.1$ figure should be read as a within-run statistical error, not a run-to-run reproducibility bound. This scatter motivates cross-checking the nested sampling evidence against an independent normalising-constant estimator, such as sequential Monte Carlo \citep{Williams2025SMC}, to establish whether it reflects a limitation of the nested sampling evidence estimate or the intrinsic difficulty of this multimodal problem; a larger ensemble of independent repeats would address the same question empirically and is left to follow-up work. The conclusion that survives this scatter is the sign-independent one -- $|\ln\mathcal{B}_{\rm B/A}|<1$ in both seeds -- so Mode~B is neither significantly favoured nor disfavoured regardless of seed. A larger seed ensemble would tighten this to a distributional rather than sign-only claim; the \texttt{analyze\_seed\_ensemble.py} script in the public data release aggregates seeds as they accumulate. Mode~B contributes a non-negligible share of the joint posterior under any prior that gives the low-\dL\ region appreciable weight. The Mode-B-anchored unrestricted run recovers $P(\hzero>120)=0.281$ -- statistically indistinguishable from the GWTC-1-anchored \IMR\ direct uniform-in-$d_L$ run reported in Section~\ref{sec:prior} -- so the heterodyne-reference choice does not bias the Mode-A/Mode-B weight ratio under \IMR/NRTidalv2; the corresponding IMRX mode-isolated set with a Mode-B-anchored reference is queued as follow-up.

This is the mechanism behind the reweighting capture fraction of equation~\ref{eq:reweight-fraction}. Mode~B is the high-\hzero, small-\dL\ branch of the distance--inclination degeneracy. The baseline volumetric prior $\pi(\dL)\propto\dL^{2}$ assigns Mode~B a tiny prior mass (the $[10,30]\,\rm Mpc$ slab carries $(30^3-10^3)/(75^3-10^3)\approx 6\,\%$ of the total $[10,75]\,\rm Mpc$ volumetric prior), and the corresponding region of the baseline posterior is sparsely sampled. Reweighting cannot reconstruct what is not there. Direct sampling under the uniform-in-$d_L$ prior populates Mode~B at sampler resolution, and it is precisely the missing Mode-B mass that opens the gap between the directly sampled and reweighted estimates of $P(\hzero>120)$. Posterior-coverage diagnostics that compare the baseline samples against the target prior -- such as the $n_{\rm eff}$ measurement reported in Section~\ref{sec:prior} -- should therefore be standard practice before reporting a reweighted bright-siren \hzero\ summary.

\subsection{Cross-waveform Hubble-constant posterior}
\label{sec:cross-waveform}

\begin{figure}
  \centering
  \includegraphics[width=\columnwidth]{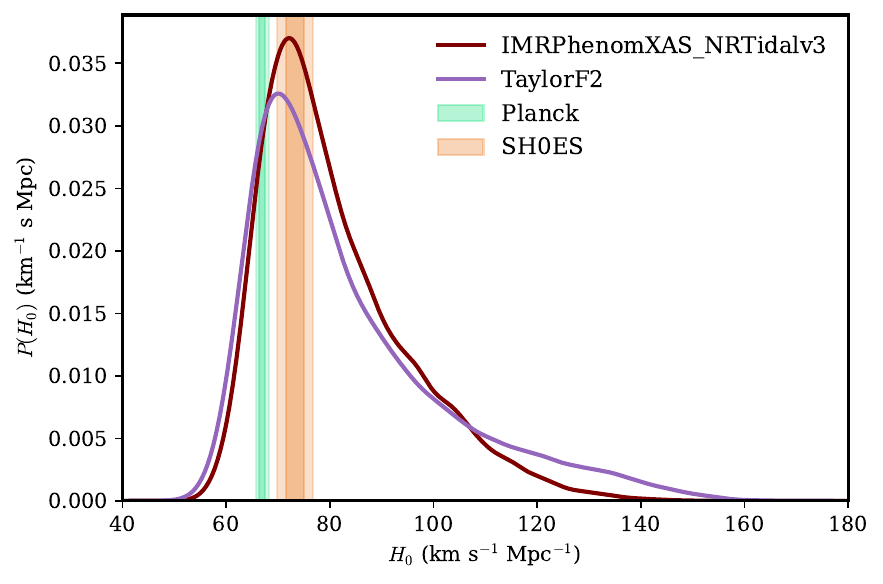}
  \caption{GW170817 \hzero\ posterior for \IMRX\ and \TF\ on the LVK-matched prior, shown as Silverman-bandwidth kernel-density estimates; the sample-derived 68 and 95 per cent HPDs (computed from the weighted samples, no KDE) are reported in Table~\ref{tab:waveform-h0}. The Planck $\Lambda$CDM and SH0ES distance ladder values are shown for context.}
  \label{fig:waveform-h0}
\end{figure}

\begin{figure*}
  \centering
  \includegraphics[width=0.92\textwidth]{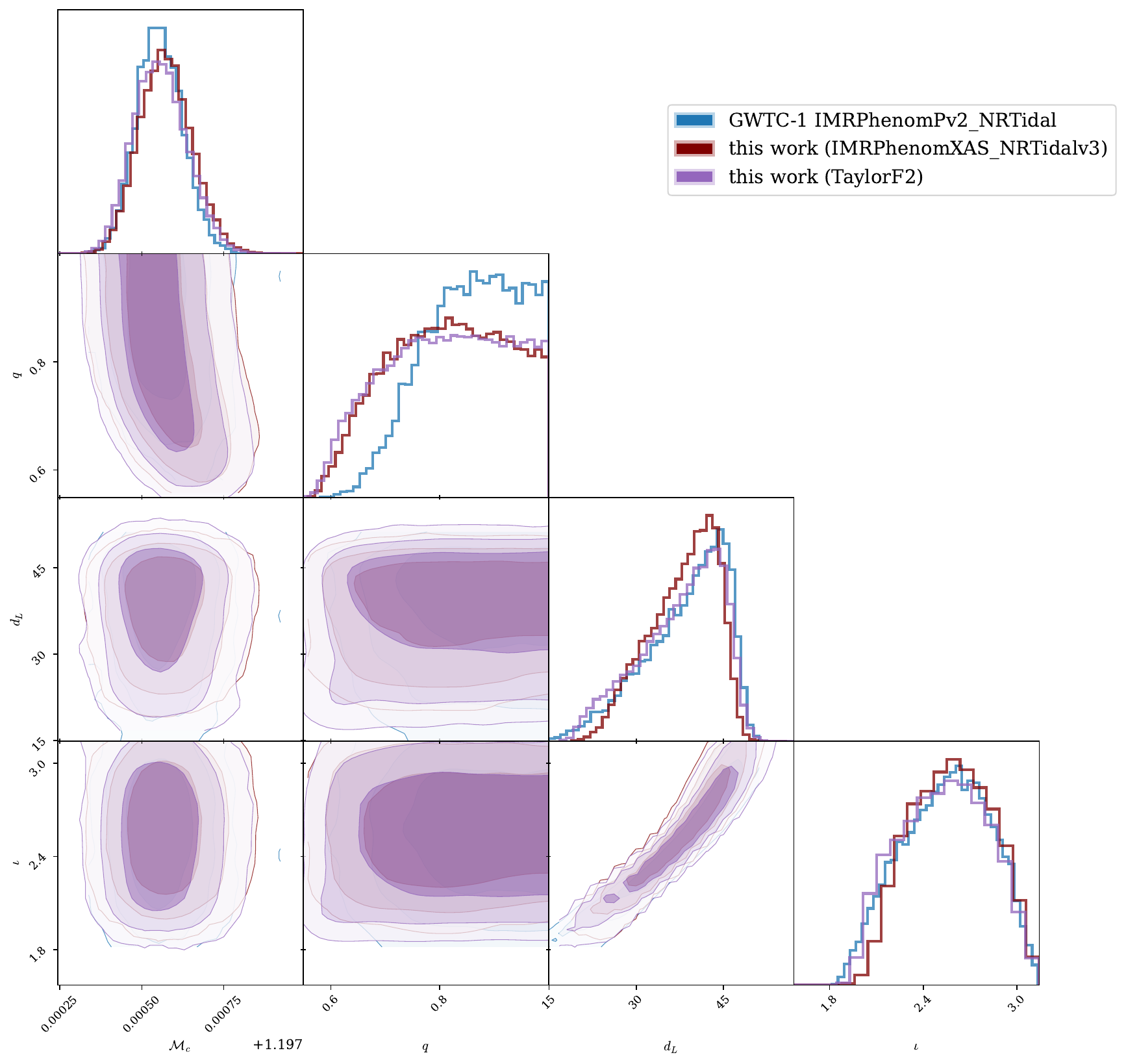}
  \caption{GW170817 source-parameter corner overlay for \IMRX\ and \TF\ on the LVK-matched prior, with the public GWTC-1 IMRPhenomPv2\_NRTidal posterior \citep{Abbott2019GWTC1,LVK_GW170817_DataRelease} included as a literature reference. Two-dimensional contours and one-dimensional marginals are KDE-smoothed.}
  \label{fig:waveform-corner}
\end{figure*}

Figure~\ref{fig:waveform-h0} shows the GW170817 \hzero\ posterior under the LVK-matched prior (Section~\ref{sec:priors}) for \IMRX\ and \TF, together with the Planck and SH0ES reference bands. The plotted curves are Silverman-bandwidth kernel-density estimates; the sample-derived HPDs (computed from the weighted samples, no KDE) are reported in Table~\ref{tab:waveform-h0}. Figure~\ref{fig:waveform-corner} shows the corresponding $(\mathcal{M}_{c},q,\chi_{\rm eff},\dL,\iota)$ corner overlay against the public GWTC-1 IMRPhenomPv2\_NRTidal posterior \citep{Abbott2019GWTC1,LVK_GW170817_DataRelease}.

\IMRX, the locked primary, has $\hzero^{\rm MAP}=70.5\kmsmpc$, 68 per cent HPD $[63.8,87.6]\kmsmpc$ and 95 per cent HPD $[59.3,112.4]\kmsmpc$; this is the baseline run of Section~\ref{sec:prior} (Table~\ref{tab:h0priors}), reused here as the reference waveform for the cross-waveform comparison. \TF, a post-Newtonian inspiral-only family check on the same prior, has $\hzero^{\rm MAP}=68.5\kmsmpc$ -- ${\sim}2\kmsmpc$ lower, the small offset expected for a lower-PN, point-particle inspiral family -- and 68 per cent HPD $[61.2,89.3]\kmsmpc$, overlapping \IMRX\ throughout. The published Abbott+2017 result $\hzero=70_{-8}^{+12}\kmsmpc$ \citep{Abbott2017H0} is consistent with both. \TF\ carries the heavier upper tail (95 per cent upper bound $127.0$ versus $112.4\kmsmpc$ for \IMRX, and $P(\hzero>120)=0.065$ versus $0.017$), which we attribute to the more recent NR-calibrated tides in \IMRX\ damping the high-\hzero, small-\dL\ branch of the degeneracy. The IMRX-vs-IMR capture-fraction comparison reported in Section~\ref{sec:prior} indicates that the same tide-driven tail compression accounts for the smaller absolute prior-induced shift in $P(\hzero>120)$ under \IMRX.

\begin{table*}
  \centering
  \caption{GW170817 \hzero\ summaries for the two-waveform main set on the LVK-matched prior, $n_{\rm live}=5000$. MAP and sample-derived HPDs in $\kmsmpc$. The \IMRX\ row is the baseline run of Table~\ref{tab:h0priors}.}
  \label{tab:waveform-h0}
  \begin{tabular}{lccccc}
    \toprule
    Waveform & MAP & 68\% HPD & 95\% HPD & $P(\hzero>120)$ & $\ln Z$ \\
    \midrule
    \IMRX                        & 70.5 & $[63.8,87.6]$ & $[59.3,112.4]$ & $0.017$ & $486.25\pm0.11$ \\
    \TF                          & 68.5 & $[61.2,89.3]$ & $[56.7,127.0]$ & $0.065$ & $486.90\pm0.10$ \\
    \bottomrule

  \end{tabular}
\end{table*}

\section{Performance and runtime studies}
\label{sec:performance}

\subsection{Wall-clock runtime}
\label{sec:wallclock}

The primary heterodyned GW170817 \hzero\ analysis with \IMRX\ and $n_{\rm live}=5000$ on the LVK-matched prior set completes in $\approx 13$\,min on a single NVIDIA A100 40\,GB SXM4 GPU (Google Cloud \texttt{a2-highgpu-1g} class); the corresponding \TF\ run takes $\approx 4$\,min. This wall-clock is comparable per live point to other JAX-based BNS pipelines \citep{Wong2023Jim,Wouters2024JimBNS} on similar hardware. The full four-variant prior-sensitivity suite (baseline, direct uniform-in-$d_L$, reweighted post-hoc, $\sigmavp=250\kms$) for \IMRX\ on the LVK-matched prior set fits inside an hour. The GW150914 \XPHM\ validation run at the larger $n_{\rm live}=8000$, $n_{\rm mcmc}=160$ settings takes $\approx 5\,\rm h$ on the same hardware -- the price for the converged short-signal sampler configuration that yields the LVK reproduction of Section~\ref{sec:gw150914} together with the \jax\ \ripple\ \XPHM\ implementation \citep{Edwards2023Ripple}, which, unlike the aligned-spin tidal waveforms used for the GW170817 science runs, is not yet performance-optimised; we expect this wall-clock to fall substantially once that implementation is optimised.

\subsection{Heterodyne speedup and live-point scaling}
\label{sec:speedup}

The heterodyned (relative-binning) likelihood is the enabling approximation: at matched live-point count it runs ${\sim}50\times$ faster than the full-resolution reference, averaged over $n_{\rm live}=500$--$2500$, with the relative-binning approximation accounting for the bulk of the reduction and the GPU saturation provided by \bjns\ layered on top. The heterodyned $\hzero^{\rm MAP}$ is stable across $n_{\rm live}\geq 2500$ to within $1\kmsmpc$ and $\ln Z$ to within $\pm 0.2$ over the same range, so the $n_{\rm live}=5000$ configuration adopted for the science runs sits in the converged regime; live-point counts up to $10^{5}$ remain accessible on a single GPU, with the $n_{\rm live}=10^{5}$ heterodyned run completing in roughly four hours. The full live-point scaling characterisation of a closely related \bjns\ nested sampling pipeline is reported by \citet{Prathaban2025GPUSpeed}; the GW170817-specific timings underlying the speedup quoted here are tabulated in \texttt{Results/scaling\_study/scaling\_summary\_full.csv} in the public data release.

\subsection{Scope of the performance claim}

The runtime numbers above are single-A100 wall-clock on the current pipeline: the full $n_{\rm live}=5000$ \hzero\ analysis completes in under a quarter of an hour, the multi-variant prior-sensitivity suite fits inside an hour, and the live-point convergence study extends to $n_{\rm live}=10^{5}$ within an overnight budget. This budget is what makes the multi-axis robustness study of this paper scientifically routine. A like-for-like \IMRX\ \pbilby\ benchmark on matched priors and live-point counts would calibrate the speedup against a production CPU pipeline and is the appropriate cross-check for a follow-up.

\section{Discussion}
\label{sec:discussion}

\subsection{Implications for bright-siren cosmology}

The main methodological message is that posterior reweighting, although statistically consistent in the limit of infinite baseline samples, can substantially understate the prior sensitivity of bright-siren \hzero\ measurements in practice. For GW170817 under \IMRX, switching from a volumetric distance prior to a uniform-in-$d_L$ prior changes $P(\hzero>120\kmsmpc)$ by a factor of $\approx 9$ under direct sampling, while reweighting captures only $\approx 17\,\%$ of that shift (equation~\ref{eq:reweight-fraction}). The corresponding capture fraction on \IMR\ -- where the high-\hzero\ tail is heavier -- rises to $\approx 58\,\%$. The deficit is a problem of effective sample coverage rather than of statistical methodology: the baseline volumetric prior assigns Mode~B a small prior mass and reweighting cannot reconstruct what was not sampled. The reweighted posterior's effective sample size, lower than the baseline's despite reweighting from the same draw, independently flags the coverage failure and is the diagnostic we recommend as a default check before any reweighted bright-siren \hzero\ summary is reported.

For a single event the cosmological interpretation is robust to this under-estimation: both early- and late-Universe \hzero\ values lie within the GW170817 68 per cent HPD under any of the priors considered here. However, the scale of the effect matters for forecasts \citep{Chen2018Forecast} and combined analyses. When bright-siren posteriors are combined over multiple events, or when tail probabilities are compared against external priors from cosmological surveys, the prior-dependence of each single-event posterior propagates through. An analysis that quantifies that dependence through reweighting alone will systematically underestimate the prior contribution.

\subsection{Where reweighting is sufficient}

Reweighting is adequate when the target and baseline priors have similar support in the regions that carry most of the posterior mass. The $\sigmavp=250\kms$ variant in our analysis is a case of this kind: the change modifies the host-velocity likelihood smoothly over the scale of the velocity posterior, the baseline samples cover the affected region well, and the reweighted and directly sampled $\sigmavp$ posteriors agree to well within the statistical width. The uniform-in-$d_L$ case fails the same coverage test, because the baseline volumetric and target uniform-in-$d_L$ priors differ most at the edges of the \dL\ range. The mode-isolated runs of Section~\ref{sec:bimodality} make this concrete: Mode~B has $|\ln \mathcal{B}_{\rm B/A}|<1$ relative to Mode~A (confirmed over two independent seeds) under the uniform-in-$d_L$ prior, but its baseline-volumetric-prior mass is only $\approx 7\,\%$ of Mode~A's, so reweighted samples are blind to it.

\subsection{Implications for future bright sirens}

The practical consequence is that direct prior sampling should be used for bright-siren \hzero\ analyses whenever the target prior differs non-trivially in support from the baseline, rather than being used only as a cross-check. The runtime budget on a single A100 GPU now permits such reruns routinely (Section~\ref{sec:performance}); the live-point and sampler-robustness studies (Appendix~\ref{app:robustness}) show that the science-quality configuration sits in the converged regime, so the marginal cost of such a rerun is the marginal cost of one ${\sim}15\,\rm min$ run. Live-point counts up to $10^{5}$ are accessible on the same hardware, which means that these studies can be combined with population-level analyses without a substantial increase in end-to-end wall-clock time. This is directly relevant to the third-generation detector era, where the detection rate of compact-binary mergers will make repeated, controlled posterior reanalyses the norm rather than the exception. Quantitatively, \citet{Chen2018Forecast} project of order $25$--$80$ bright-siren events with electromagnetic counterparts within 5--10 years from a network including A+ LIGO and Voyager, with per-event posterior widths comparable to GW170817's; for the Einstein Telescope and Cosmic Explorer era, the projected compact-binary detection rate brings the bright-siren sub-population to of order tens to hundreds per year \citep[e.g.][]{Branchesi2023ET}, a regime in which the per-event parameter-estimation cost itself becomes a bottleneck \citep{HuVeitch2025}. At the $\lesssim 15\,\rm min$-per-event runtime reported in Section~\ref{sec:performance} this makes per-event prior-sensitivity reruns routine across that population. For comparison-grade single-event posteriors we recommend reporting both the volumetric (which coincides with the comoving-volume prior at $z\lesssim 0.02$) and the uniform-in-$d_L$ direct-sampled posteriors; the difference between the two bounds the prior-induced systematic.

\section{Conclusions}
\label{sec:conclusions}

We have used a GPU-native nested sampling pipeline to revisit the GW170817 bright-siren \hzero\ measurement of \citet{Abbott2017H0}. Switching the distance prior from volumetric to uniform-in-$d_L$ by direct sampling raises $P(\hzero>120\kmsmpc)$ from $0.017$ to $0.159$ and moves the weighted-median \hzero\ from $77.6$ to $87.6\kmsmpc$, while the binned MAP stays at $70.5\kmsmpc$; post-hoc reweighting captures only $0.041$ in the tail, $\approx 17\,\%$ of the directly sampled shift. The three \IMRX\ prior variants that carry an independent evidence agree to $\Delta\ln Z\lesssim 1.8$, so the tail and bulk shifts are properties of the prior, not data updates. The mechanism is the $\dL$--$\iota$ bimodality: a high-\hzero, small-\dL\ branch (Mode~B; $|\ln\mathcal{B}_{\rm B/A}|<1$, confirmed in two independent seeds) carries appreciable likelihood but negligible volumetric-prior mass, and reweighting cannot recover what was not sampled. The reweighted estimator's effective sample size, lower than the baseline's, independently flags the coverage failure. The runtime budget makes direct prior-sensitivity reruns the right default for bright-siren cosmology, with multi-axis robustness studies now routine on a single A100 GPU. Methodologically, the runtime budget delivered by the GPU-native Nested Slice Sampling pipeline is what turns a systematic, multi-axis prior-sensitivity study from a major undertaking into a routine one, and we expect fast differentiable pipelines of this kind to make such studies standard in bright-siren cosmology.

Two natural extensions sit outside the scope of the present analysis. First, no waveform in our \jax\ inventory simultaneously includes precession and tides, so the locked primary is the aligned-spin tidal \IMRX; a tides-with-precession waveform in \jax\ would extend the cross-waveform check. Second, the $\dL$--$\iota$ bimodality is characterised here at the level of local evidences and mode-isolated marginals; propagating Mode~B into a population analysis, or combining it with an electromagnetic likelihood for the host inclination, is the natural next step for bright-siren cosmology built on this finding.

\section*{Acknowledgements}

This work was supported by the research environment and infrastructure of the Handley Lab at the University of Cambridge. This material is based upon work supported by the Google Cloud research credits program, with the award GCP397499138. We acknowledge the LIGO--Virgo--KAGRA Collaboration for the public strain data and reference posteriors used here \citep{Abbott2019GWTC1,Abbott2021GWTC2p1,LVK_GW170817_DataRelease,LVK_H0_DataRelease,GWTC2p1_GW150914_Zenodo}.

\section*{Data Availability}

The GW170817 strain data, host-galaxy peculiar-velocity inputs, and GWTC-1 reference posteriors used here are from the LIGO--Virgo Collaboration data releases LIGO P1800061 and P1700296 \citep{LVK_GW170817_DataRelease,LVK_H0_DataRelease}. The GW150914 strain data and the GWTC-2.1 \XPHM\ reference parameter-estimation posterior are from the LVK GWTC-2.1 Zenodo deposit \citep{GWTC2p1_GW150914_Zenodo}. The analysis scripts, derived summary tables, run catalogue, and figure- and table-generation code needed to reproduce the results in this paper can be obtained from \citet{Yang2026DataRelease}. The nested sampling chains themselves are not redistributed in that repository (each individual chain is ${\sim}100\,\rm MB$ and the full set runs to several GB); they can be regenerated from the public strain data using the \bjns\ nested slice sampling kernel \citep{Yallup2026BlackjaxNS}; the broader GW-analysis tooling and example scripts of \citet{Prathaban2025GPUSpeed} provide instructions for running GW analyses of this kind.

\bibliographystyle{mnras}
\bibliography{references}

\begin{thebibliography}{}
\makeatletter
\relax
\def\mn@urlcharsother{\let\do\@makeother \do\$\do\&\do\#\do\^\do\_\do\%\do\~}
\def\mn@doi{\begingroup\mn@urlcharsother \@ifnextchar [ {\mn@doi@}
  {\mn@doi@[]}}
\def\mn@doi@[#1]#2{\def\@tempa{#1}\ifx\@tempa\@empty \href
  {http://dx.doi.org/#2} {doi:#2}\else \href {http://dx.doi.org/#2} {#1}\fi
  \endgroup}
\def\mn@eprint#1#2{\mn@eprint@#1:#2::\@nil}
\def\mn@eprint@arXiv#1{\href {http://arxiv.org/abs/#1} {{\tt arXiv:#1}}}
\def\mn@eprint@dblp#1{\href {http://dblp.uni-trier.de/rec/bibtex/#1.xml}
  {dblp:#1}}
\def\mn@eprint@#1:#2:#3:#4\@nil{\def\@tempa {#1}\def\@tempb {#2}\def\@tempc
  {#3}\ifx \@tempc \@empty \let \@tempc \@tempb \let \@tempb \@tempa \fi \ifx
  \@tempb \@empty \def\@tempb {arXiv}\fi \@ifundefined
  {mn@eprint@\@tempb}{\@tempb:\@tempc}{\expandafter \expandafter \csname
  mn@eprint@\@tempb\endcsname \expandafter{\@tempc}}}

\bibitem[\protect\citeauthoryear{{Abac}, {Dietrich}, {Buonanno}, {Steinhoff}
  \& {Ujevic}}{{Abac} et~al.}{2024}]{Abac2024NRTidalv3}
{Abac} A.,  {Dietrich} T.,  {Buonanno} A.,  {Steinhoff} J.,   {Ujevic} M.,
  2024, \mn@doi [Physical Review D] {10.1103/PhysRevD.109.024062}, 109, 024062

\bibitem[\protect\citeauthoryear{{Abbott} et~al.}{{Abbott}
  et~al.}{2016}]{Abbott2016GW150914}
{Abbott} B.~P.,  et~al., 2016, \mn@doi [Physical Review Letters]
  {10.1103/PhysRevLett.116.061102}, 116, 061102

\bibitem[\protect\citeauthoryear{{Abbott} et~al.}{{Abbott}
  et~al.}{2017a}]{Abbott2017GW170817Discovery}
{Abbott} B.~P.,  et~al., 2017a, \mn@doi [Physical Review Letters]
  {10.1103/PhysRevLett.119.161101}, 119, 161101

\bibitem[\protect\citeauthoryear{{Abbott} et~al.}{{Abbott}
  et~al.}{2017b}]{Abbott2017H0}
{Abbott} B.~P.,  et~al., 2017b, \mn@doi [Nature] {10.1038/nature24471}, 551, 85

\bibitem[\protect\citeauthoryear{{Abbott} et~al.}{{Abbott}
  et~al.}{2019a}]{Abbott2019GW170817Properties}
{Abbott} B.~P.,  et~al., 2019a, \mn@doi [Physical Review X]
  {10.1103/PhysRevX.9.011001}, 9, 011001

\bibitem[\protect\citeauthoryear{{Abbott} et~al.}{{Abbott}
  et~al.}{2019b}]{Abbott2019GWTC1}
{Abbott} B.~P.,  et~al., 2019b, \mn@doi [Physical Review X]
  {10.1103/PhysRevX.9.031040}, 9, 031040

\bibitem[\protect\citeauthoryear{{Abbott} et~al.}{{Abbott}
  et~al.}{2020}]{LVK2020Prospects}
{Abbott} B.~P.,  et~al., 2020, \mn@doi [Living Reviews in Relativity]
  {10.1007/s41114-020-00026-9}, 23, 3

\bibitem[\protect\citeauthoryear{{Abbott} et~al.}{{Abbott}
  et~al.}{2024}]{Abbott2021GWTC2p1}
{Abbott} R.,  et~al., 2024, \mn@doi [Physical Review D]
  {10.1103/PhysRevD.109.022001}, 109, 022001

\bibitem[\protect\citeauthoryear{{Ashton}}{{Ashton}}{2026}]{Ashton2025Reweighting}
{Ashton} G.,  2026, \mn@doi [RAS Techniques and Instruments]
  {10.1093/rasti/rzag012}, 5, rzag012

\bibitem[\protect\citeauthoryear{{Ashton} et~al.}{{Ashton}
  et~al.}{2019}]{Ashton2019Bilby}
{Ashton} G.,  et~al., 2019, \mn@doi [The Astrophysical Journal Supplement
  Series] {10.3847/1538-4365/ab06fc}, 241, 27

\bibitem[\protect\citeauthoryear{{Bradbury} et~al.}{{Bradbury}
  et~al.}{2018}]{Bradbury2018JAX}
{Bradbury} J.,  et~al., 2018, {JAX: composable transformations of Python+NumPy
  programs}, \url{https://github.com/jax-ml/jax}

\bibitem[\protect\citeauthoryear{{Branchesi} et~al.}{{Branchesi}
  et~al.}{2023}]{Branchesi2023ET}
{Branchesi} M.,  et~al., 2023, \mn@doi [Journal of Cosmology and Astroparticle
  Physics] {10.1088/1475-7516/2023/07/068}, 2023, 068

\bibitem[\protect\citeauthoryear{{Cabezas}, {Corenflos}, {Lao}, {Louf}
  et~al.}{{Cabezas} et~al.}{2024}]{BlackJAX}
{Cabezas} A.,  {Corenflos} A.,  {Lao} J.,  {Louf} R.,   et~al., 2024,
  {BlackJAX: Composable Bayesian inference in JAX} (\mn@eprint {arXiv}
  {2402.10797})

\bibitem[\protect\citeauthoryear{{Chan}, {Narola}, {Ng}, {Wouters}, {Wong},
  {Gittins}, {Janquart}  \& {Van Den Broeck}}{{Chan}
  et~al.}{2026}]{Chan2026ripple}
{Chan} R.,  {Narola} H.,  {Ng} T. C.~K.,  {Wouters} T.,  {Wong} I. C.~F.,
  {Gittins} F.,  {Janquart} J.,   {Van Den Broeck} C.,  2026, in prep.

\bibitem[\protect\citeauthoryear{{Chen}, {Fishbach}  \& {Holz}}{{Chen}
  et~al.}{2018}]{Chen2018Forecast}
{Chen} H.-Y.,  {Fishbach} M.,   {Holz} D.~E.,  2018, \mn@doi [Nature]
  {10.1038/s41586-018-0606-0}, 562, 545

\bibitem[\protect\citeauthoryear{{Cornish}}{{Cornish}}{2010}]{Cornish2010}
{Cornish} N.~J.,  2010, {Fast Fisher Matrices and Lazy Likelihoods} (\mn@eprint
  {arXiv} {1007.4820})

\bibitem[\protect\citeauthoryear{{Dax} et~al.,}{{Dax}
  et~al.}{2025}]{Dax2025DingoBNS}
{Dax} M.,  et~al., 2025, \mn@doi [Nature] {10.1038/s41586-025-08593-z}, 639, 49

\bibitem[\protect\citeauthoryear{{Di Valentino} et~al.}{{Di Valentino}
  et~al.}{2021}]{DiValentino2021HubbleReview}
{Di Valentino} E.,  et~al., 2021, \mn@doi [Classical and Quantum Gravity]
  {10.1088/1361-6382/ac086d}, 38, 153001

\bibitem[\protect\citeauthoryear{{Edwards}, {Wong}, {Lam}, {Coogan},
  {Foreman-Mackey}, {Isi}  \& {Zimmerman}}{{Edwards}
  et~al.}{2024}]{Edwards2023Ripple}
{Edwards} T. D.~P.,  {Wong} K. W.~K.,  {Lam} K. K.~H.,  {Coogan} A.,
  {Foreman-Mackey} D.,  {Isi} M.,   {Zimmerman} A.,  2024, \mn@doi [Physical
  Review D] {10.1103/PhysRevD.110.064028}, 110, 064028

\bibitem[\protect\citeauthoryear{{Finn} \& {Chernoff}}{{Finn} \&
  {Chernoff}}{1993}]{FinnChernoff1993}
{Finn} L.~S.,  {Chernoff} D.~F.,  1993, \mn@doi [Phys. Rev. D]
  {10.1103/PhysRevD.47.2198}, 47, 2198

\bibitem[\protect\citeauthoryear{{Holz} \& {Hughes}}{{Holz} \&
  {Hughes}}{2005}]{HolzHughes2005}
{Holz} D.~E.,  {Hughes} S.~A.,  2005, \mn@doi [The Astrophysical Journal]
  {10.1086/431341}, 629, 15

\bibitem[\protect\citeauthoryear{{Hotokezaka}, {Nakar}, {Gottlieb}, {Nissanke},
  {Masuda}, {Hallinan}, {Mooley}  \& {Deller}}{{Hotokezaka}
  et~al.}{2019}]{Hotokezaka2019}
{Hotokezaka} K.,  {Nakar} E.,  {Gottlieb} O.,  {Nissanke} S.,  {Masuda} K.,
  {Hallinan} G.,  {Mooley} K.~P.,   {Deller} A.~T.,  2019, \mn@doi [Nature
  Astronomy] {10.1038/s41550-019-0820-1}, 3, 940

\bibitem[\protect\citeauthoryear{{Howlett} \& {Davis}}{{Howlett} \&
  {Davis}}{2020}]{HowlettDavis2020}
{Howlett} C.,  {Davis} T.~M.,  2020, \mn@doi [Monthly Notices of the Royal
  Astronomical Society] {10.1093/mnras/staa049}, 492, 3803

\bibitem[\protect\citeauthoryear{{Hu} \& {Veitch}}{{Hu} \&
  {Veitch}}{2025}]{HuVeitch2025}
{Hu} Q.,  {Veitch} J.,  2025, \mn@doi [Physical Review D] {10.1103/dj7k-tk37},
  112, 084039

\bibitem[\protect\citeauthoryear{{Kish}}{{Kish}}{1965}]{Kish1965}
{Kish} L.,  1965, {Survey Sampling}.
John Wiley \& Sons, New York

\bibitem[\protect\citeauthoryear{{Krishna}, {Vijaykumar}, {Ganguly}, {Talbot},
  {Biscoveanu}, {George}, {Williams}  \& {Zimmerman}}{{Krishna}
  et~al.}{2023}]{Krishna2023RelativeBinningBilby}
{Krishna} K.,  {Vijaykumar} A.,  {Ganguly} A.,  {Talbot} C.,  {Biscoveanu} S.,
  {George} R.~N.,  {Williams} N.,   {Zimmerman} A.,  2023, {Accelerated
  parameter estimation in Bilby with relative binning} (\mn@eprint {arXiv}
  {2312.06009})

\bibitem[\protect\citeauthoryear{{LIGO Scientific Collaboration} \& {Virgo
  Collaboration}}{{LIGO Scientific Collaboration} \& {Virgo
  Collaboration}}{2017}]{LVK_H0_DataRelease}
{LIGO Scientific Collaboration} {Virgo Collaboration} 2017, {A
  gravitational-wave standard siren measurement of the Hubble constant -- Data
  Release}, LIGO Document P1700296;
  \url{https://dcc.ligo.org/LIGO-P1700296/public}

\bibitem[\protect\citeauthoryear{{LIGO Scientific Collaboration} \& {Virgo
  Collaboration}}{{LIGO Scientific Collaboration} \& {Virgo
  Collaboration}}{2018}]{LVK_GW170817_DataRelease}
{LIGO Scientific Collaboration} {Virgo Collaboration} 2018, {Properties of the
  binary neutron star merger GW170817 -- Data Release}, LIGO Document P1800061;
  \url{https://dcc.ligo.org/LIGO-P1800061/public}

\bibitem[\protect\citeauthoryear{{LIGO Scientific Collaboration} \& {Virgo
  Collaboration}}{{LIGO Scientific Collaboration} \& {Virgo
  Collaboration}}{2022}]{GWTC2p1_GW150914_Zenodo}
{LIGO Scientific Collaboration} {Virgo Collaboration} 2022, {GWTC-2.1: Deep
  Extended Catalog of Compact Binary Coalescences Observed by LIGO and Virgo
  During the First Half of the Third Observing Run -- Parameter Estimation Data
  Release}, Zenodo, \url{https://doi.org/10.5281/zenodo.6513631},
  \mn@doi{10.5281/zenodo.6513631}

\bibitem[\protect\citeauthoryear{{LIGO Scientific Collaboration}, {Virgo
  Collaboration}  \& {KAGRA Collaboration}}{{LIGO Scientific Collaboration}
  et~al.}{2026}]{LVK2026GWTC5Cosmology}
{LIGO Scientific Collaboration} {Virgo Collaboration}  {KAGRA Collaboration}
  2026, {GWTC-5.0: Constraints on the Cosmic Expansion Rate and Modified
  Gravitational-wave Propagation} (\mn@eprint {arXiv} {2605.27227}),
  \mn@doi{10.48550/arXiv.2605.27227}

\bibitem[\protect\citeauthoryear{{Mooley} et~al.,}{{Mooley}
  et~al.}{2018}]{Mooley2018Nature}
{Mooley} K.~P.,  et~al., 2018, \mn@doi [Nature] {10.1038/s41586-018-0486-3},
  561, 355

\bibitem[\protect\citeauthoryear{{Morisaki}}{{Morisaki}}{2021}]{Morisaki2021Multiband}
{Morisaki} S.,  2021, \mn@doi [Physical Review D]
  {10.1103/PhysRevD.104.044062}, 104, 044062

\bibitem[\protect\citeauthoryear{{Mukherjee}, {Lavaux}, {Bouchet}, {Jasche},
  {Wandelt}, {Nissanke}, {Leclercq}  \& {Hotokezaka}}{{Mukherjee}
  et~al.}{2021}]{Mukherjee2021Velocity}
{Mukherjee} S.,  {Lavaux} G.,  {Bouchet} F.~R.,  {Jasche} J.,  {Wandelt} B.~D.,
   {Nissanke} S.,  {Leclercq} F.,   {Hotokezaka} K.,  2021, \mn@doi [Astronomy
  and Astrophysics] {10.1051/0004-6361/201936724}, 646, A65

\bibitem[\protect\citeauthoryear{{Nicolaou}, {Lahav}, {Lemos}, {Hartley}  \&
  {Braden}}{{Nicolaou} et~al.}{2020}]{Nicolaou2020}
{Nicolaou} C.,  {Lahav} O.,  {Lemos} P.,  {Hartley} W.,   {Braden} J.,  2020,
  \mn@doi [Monthly Notices of the Royal Astronomical Society]
  {10.1093/mnras/staa1120}, 495, 90

\bibitem[\protect\citeauthoryear{{Owen}}{{Owen}}{2013}]{Owen2013MonteCarlo}
{Owen} A.~B.,  2013, {Monte Carlo Theory, Methods and Examples},
  \url{https://artowen.su.domains/mc/}

\bibitem[\protect\citeauthoryear{{Palmese} \& {Mastrogiovanni}}{{Palmese} \&
  {Mastrogiovanni}}{2025}]{PalmeseMastrogiovanni2025}
{Palmese} A.,  {Mastrogiovanni} S.,  2025, {Gravitational Wave Cosmology}
  (\mn@eprint {arXiv} {2502.00239})

\bibitem[\protect\citeauthoryear{{Palmese}, {Kaur}, {Hajela}, {Margutti},
  {McDowell}  \& {MacFadyen}}{{Palmese} et~al.}{2024}]{Palmese2024GW170817H0}
{Palmese} A.,  {Kaur} R.,  {Hajela} A.,  {Margutti} R.,  {McDowell} A.,
  {MacFadyen} A.,  2024, \mn@doi [Physical Review D]
  {10.1103/PhysRevD.109.063508}, 109, 063508

\bibitem[\protect\citeauthoryear{{Payne}, {Talbot}  \& {Thrane}}{{Payne}
  et~al.}{2019}]{Payne2019Reweighting}
{Payne} E.,  {Talbot} C.,   {Thrane} E.,  2019, \mn@doi [Physical Review D]
  {10.1103/PhysRevD.100.123017}, 100, 123017

\bibitem[\protect\citeauthoryear{{Planck Collaboration}}{{Planck
  Collaboration}}{2020}]{Planck2020}
{Planck Collaboration} 2020, \mn@doi [Astronomy and Astrophysics]
  {10.1051/0004-6361/201833910}, 641, A6

\bibitem[\protect\citeauthoryear{{Prathaban}, {Yallup}, {Alvey}, {Yang},
  {Templeton}  \& {Handley}}{{Prathaban} et~al.}{2026}]{Prathaban2025GPUSpeed}
{Prathaban} M.,  {Yallup} D.,  {Alvey} J.,  {Yang} M.~H.,  {Templeton} W.,
  {Handley} W.,  2026, \mn@doi [RAS Techniques and Instruments]
  {10.1093/rasti/rzag034}, 5, rzag034

\bibitem[\protect\citeauthoryear{{Pratten}, {Husa}, {Garc\'ia-Quir\'os},
  {Colleoni}, {Ramos-Buades}, {Estell\'es}  \& {Jaume}}{{Pratten}
  et~al.}{2020}]{Pratten2020XAS}
{Pratten} G.,  {Husa} S.,  {Garc\'ia-Quir\'os} C.,  {Colleoni} M.,
  {Ramos-Buades} A.,  {Estell\'es} H.,   {Jaume} R.,  2020, \mn@doi [Physical
  Review D] {10.1103/PhysRevD.102.064001}, 102, 064001

\bibitem[\protect\citeauthoryear{{Pratten} et~al.}{{Pratten}
  et~al.}{2021}]{Pratten2021XPHM}
{Pratten} G.,  et~al., 2021, \mn@doi [Physical Review D]
  {10.1103/PhysRevD.103.104056}, 103, 104056

\bibitem[\protect\citeauthoryear{{Riess} et~al.}{{Riess}
  et~al.}{2016}]{Riess2016}
{Riess} A.~G.,  et~al., 2016, \mn@doi [The Astrophysical Journal]
  {10.3847/0004-637X/826/1/56}, 826, 56

\bibitem[\protect\citeauthoryear{{Riess} et~al.}{{Riess}
  et~al.}{2022}]{Riess2022}
{Riess} A.~G.,  et~al., 2022, \mn@doi [The Astrophysical Journal Letters]
  {10.3847/2041-8213/ac5c5b}, 934, L7

\bibitem[\protect\citeauthoryear{{Salvarese} \& {Chen}}{{Salvarese} \&
  {Chen}}{2024}]{SalvareseChen2024}
{Salvarese} A.,  {Chen} H.-Y.,  2024, \mn@doi [The Astrophysical Journal
  Letters] {10.3847/2041-8213/ad7bbc}, 974, L16

\bibitem[\protect\citeauthoryear{{Schutz}}{{Schutz}}{1986}]{Schutz1986}
{Schutz} B.~F.,  1986, \mn@doi [Nature] {10.1038/323310a0}, 323, 310

\bibitem[\protect\citeauthoryear{{Skilling}}{{Skilling}}{2006}]{Skilling2006Nested}
{Skilling} J.,  2006, \mn@doi [Bayesian Analysis] {10.1214/06-BA127}, 1, 833

\bibitem[\protect\citeauthoryear{{Smith}, {Field}, {Blackburn}, {Haster},
  {P\"urrer}, {Raymond}  \& {Schmidt}}{{Smith} et~al.}{2016}]{Smith2016ROQ}
{Smith} R.,  {Field} S.~E.,  {Blackburn} K.,  {Haster} C.-J.,  {P\"urrer} M.,
  {Raymond} V.,   {Schmidt} P.,  2016, \mn@doi [Physical Review D]
  {10.1103/PhysRevD.94.044031}, 94, 044031

\bibitem[\protect\citeauthoryear{{Speagle}}{{Speagle}}{2020}]{Speagle2020Dynesty}
{Speagle} J.~S.,  2020, \mn@doi [Monthly Notices of the Royal Astronomical
  Society] {10.1093/mnras/staa278}, 493, 3132

\bibitem[\protect\citeauthoryear{{Vehtari}, {Simpson}, {Gelman}, {Yao}  \&
  {Gabry}}{{Vehtari} et~al.}{2024}]{Vehtari2024PSIS}
{Vehtari} A.,  {Simpson} D.,  {Gelman} A.,  {Yao} Y.,   {Gabry} J.,  2024,
  Journal of Machine Learning Research, 25, 1

\bibitem[\protect\citeauthoryear{{Vinciguerra}, {Veitch}  \&
  {Mandel}}{{Vinciguerra} et~al.}{2017}]{Vinciguerra2017Multiband}
{Vinciguerra} S.,  {Veitch} J.,   {Mandel} I.,  2017, \mn@doi [Classical and
  Quantum Gravity] {10.1088/1361-6382/aa6d44}, 34, 115006

\bibitem[\protect\citeauthoryear{{Wang}, {Tang}, {Jin}  \& {Fan}}{{Wang}
  et~al.}{2023}]{Wang2023GW170817H0}
{Wang} Y.-Y.,  {Tang} S.-P.,  {Jin} Z.-P.,   {Fan} Y.-Z.,  2023, \mn@doi [The
  Astrophysical Journal] {10.3847/1538-4357/aca96c}, 943, 13

\bibitem[\protect\citeauthoryear{{Williams}, {Veitch}  \&
  {Messenger}}{{Williams} et~al.}{2021}]{Williams2021Nessai}
{Williams} M.~J.,  {Veitch} J.,   {Messenger} C.,  2021, \mn@doi [Physical
  Review D] {10.1103/PhysRevD.103.103006}, 103, 103006

\bibitem[\protect\citeauthoryear{{Williams}, {Karamanis}, {Luo}  \&
  {Seljak}}{{Williams} et~al.}{2025}]{Williams2025SMC}
{Williams} M.~J.,  {Karamanis} M.,  {Luo} Y.,   {Seljak} U.,  2025, \mn@doi
  [Monthly Notices of the Royal Astronomical Society] {10.1093/mnras/staf1458},
  543, 1479

\bibitem[\protect\citeauthoryear{{Wong}, {Isi}  \& {Edwards}}{{Wong}
  et~al.}{2023}]{Wong2023Jim}
{Wong} K. W.~K.,  {Isi} M.,   {Edwards} T. D.~P.,  2023, \mn@doi [The
  Astrophysical Journal] {10.3847/1538-4357/acf5cd}, 958, 129

\bibitem[\protect\citeauthoryear{{Wouters}, {Pang}, {Dietrich}  \& {Van Den
  Broeck}}{{Wouters} et~al.}{2024}]{Wouters2024JimBNS}
{Wouters} T.,  {Pang} P. T.~H.,  {Dietrich} T.,   {Van Den Broeck} C.,  2024,
  \mn@doi [Physical Review D] {10.1103/PhysRevD.110.083033}, 110, 083033

\bibitem[\protect\citeauthoryear{{Yallup}, {Prathaban}, {Alvey}  \&
  {Handley}}{{Yallup} et~al.}{2025}]{Yallup2025BlackJAXNS}
{Yallup} D.,  {Prathaban} M.,  {Alvey} J.,   {Handley} W.,  2025, {Parallel
  Nested Slice Sampling for Gravitational Wave Parameter Estimation}
  (\mn@eprint {arXiv} {2509.24949})

\bibitem[\protect\citeauthoryear{{Yallup}, {Kroupa}  \& {Handley}}{{Yallup}
  et~al.}{2026}]{Yallup2026BlackjaxNS}
{Yallup} D.,  {Kroupa} N.,   {Handley} W.,  2026, \mn@doi [Transactions on
  Machine Learning Research] {10.48550/arXiv.2601.23252}

\bibitem[\protect\citeauthoryear{{Yang}, {Prathaban}, {Yallup}  \&
  {Handley}}{{Yang} et~al.}{2026}]{Yang2026DataRelease}
{Yang} M.~H.,  {Prathaban} M.,  {Yallup} D.,   {Handley} W.,  2026, {GW170817
  bright-siren $H_0$: data and analysis release},
  \url{https://github.com/ming-256/GW170817-bright-siren-H0},
  \mn@doi{10.5281/zenodo.21038511}

\bibitem[\protect\citeauthoryear{{Zackay}, {Dai}  \& {Venumadhav}}{{Zackay}
  et~al.}{2018}]{Zackay2018RelativeBinning}
{Zackay} B.,  {Dai} L.,   {Venumadhav} T.,  2018, {Relative binning and fast
  likelihood evaluation for gravitational-wave parameter estimation}
  (\mn@eprint {arXiv} {1806.08792})

\makeatother
\end{thebibliography}

\appendix

\section{Inline robustness sweeps}
\label{app:robustness}

We complete the multi-axis robustness picture with sweeps held in this appendix; the prior-sensitivity result of Section~\ref{sec:prior} survives every one of them.

\emph{Sampler hyperparameters.} On the \IMR\ LVK-matched baseline at $n_{\rm live}=5000$, varying $n_{\rm delete}/n_{\rm live}\in\{0.10,0.25,0.50,0.75\}$ moves $\hzero^{\rm MAP}$ by $\le 1.2\kmsmpc$ and $\ln Z$ by $\le 0.6$; varying the heterodyne-bin count $n_{\rm bins}\in\{251,501,1001\}$ moves $\hzero^{\rm MAP}$ by $\le 0.7\kmsmpc$, with the largest pairwise Wasserstein-1 distance between the three \hzero\ posteriors below $2\kmsmpc$.

\emph{PSD source.} \TF\ runs at $n_{\rm live}=5000$ on the baseline prior with GWTC-1, \texttt{kazewong}, and \bilby\ reference PSDs span $\hzero^{\rm MAP}=76.4$--$77.7\kmsmpc$ and $P(\hzero>120)=0.043$--$0.065$, again with maximum pairwise W$_{1}<2\kmsmpc$.

\emph{Heterodyne reference.} Replacing the GWTC-1 reference parameters with an optimised (maximum-likelihood) reference point from an initial low-cost optimisation produces \IMR\ posteriors that are indistinguishable through the bulk of the distribution and depart only in the high-\hzero\ tail.

\emph{Peculiar-velocity centre.} Sweeping $\langle v_{p}\rangle\in\{215, 310, 405\}\kms$ at fixed $\sigmavp=150\kms$ on the \IMRX\ baseline (Figure~\ref{fig:h0prior}b, Table~\ref{tab:h0priors}) shifts $\hzero^{\rm MAP}$ by $6\kmsmpc$ peak-to-peak across the full range ($74.5$ at $\langle v_{p}\rangle=215\kms$ to $68.5$ at $\langle v_{p}\rangle=405\kms$), while $P(\hzero>120)$ changes by less than $0.02$. The $\langle v_{p}\rangle=310\kms$ run reproduces the \IMRX\ baseline to within $0.01$ in $\ln Z$, confirming run reproducibility.

\emph{\IMR\ companion full sweep.} The four-variant prior-sensitivity sweep on \IMR\ (Table~\ref{tab:imr-sweep}) reproduces the \IMRX\ behaviour with a larger amplitude: the reweighting-capture fraction is $(0.195-0.076)/(0.281-0.076)\approx 58\,\%$, against $\approx 17\,\%$ for \IMRX\ (equation~\ref{eq:reweight-fraction}).

\begin{table}
  \centering
  \caption{\IMR\ companion four-variant prior-sensitivity sweep, the legacy-waveform analogue of Table~\ref{tab:h0priors}. MAP in $\kmsmpc$.}
  \label{tab:imr-sweep}
  \begin{tabular}{lcc}
    \toprule
    Variant & $\hzero^{\rm MAP}$ & $P(\hzero>120)$ \\
    \midrule
    Volumetric baseline           & $71.5$ & $0.076$ \\
    Uniform-in-$d_L$ (direct)     & $73.5$ & $0.281$ \\
    Uniform-in-$d_L$ (reweighted) & $71.5$ & $0.195$ \\
    $\sigmavp=250\kms$            & $70.5$ & $0.067$ \\
    \bottomrule
  \end{tabular}
\end{table}

All five axes therefore produce \hzero\ shifts at least an order of magnitude smaller than the prior-induced shift documented in Section~\ref{sec:prior}.

\label{lastpage}
\end{document}